\documentclass[twocolumn,showpacs,preprintnumbers,amsmath,amssymb]{revtex4}
\usepackage{amsmath,amssymb,bm,graphicx,dcolumn}
\numberwithin{equation}{section}

\begin{document}
\title{Theory of preparation and relaxation of a \textit{p}-orbital atomic Mott insulator}

\author{John H. Challis}
 \email{john.challis@yale.edu}
\author{S.M. Girvin}%
\author{L.I. Glazman}
\affiliation{Department
  of Physics, Yale University, New Haven, CT 06520}

\date{\today}

\begin{abstract}
We develop a theoretical framework to understand the preparation and relaxation of a metastable Mott insulator state  within the first excited band of a 1D optical
  lattice. The state is loaded by ``lifting" atoms from the ground to
  the first excited band by means of a stimulated Raman transition. We determine the
  effect of pulse duration on the accuracy of the state preparation for the case of a Gaussian pulse shape. Relaxation of the prepared state occurs in two major stages: double-occupied sites occurring due to quantum fluctuations initially lead to interband transitions followed by a spreading of particles in the trap and thermalization. We find the characteristic relaxation times at the earliest stage and at asymptotically long times approaching equilibrium. Our theory is applicable to recent experiments performed with 1D optical lattices [T. M\"uller, S. F\"olling, A. Widera, and I. Bloch, Phys. Rev. Lett. \textbf{99}, 200405 (2007)].
\end{abstract}
\maketitle

\section{\label{sec:intro}Introduction}
Cold atoms loaded into an optical lattice are a natural system to study many-body physics far from equilibrium \cite{Bloch2007}. Experiments at the University of Mainz and NIST have succeeded in promoting a large fraction of bosons from the ground band to excited lattice orbitals \cite{Muller2007,Spielman2006}. Many-body phenomena within these excited orbitals can be strikingly different from the ground band; theoretical proposals include the existence of supersolid, algebraic bond liquid, and $p$-wave superfluid phases \cite{Scarola2005,Isacsson05,Xucondmat2007,Kuklov2006,Xu2007,Wu2006,Wu2007,Wu2006_2,Wu2009}. The inherent metastability of the excited band is the main experimental barrier to realizing these models in optical lattice systems. Isacsson et al. pointed out that the absence of phase space for initial decay can lead to long excited band lifetimes $1/\Gamma$ when compared to the nearest neighbor hopping time \cite{Isacsson05}. M\"uller et al. confirmed this prediction experimentally \cite{Muller2007}. 
In a related system, Spielman et al. studied the preparation and relaxation of bosons promoted to the first excited band in an array of quasi-2D lattices \cite{Spielman2006}. Recently, Stojanovi\'c et al. calculated the excited band lifetime for a superfluid of bosons in a shallow 2D double-well optical lattice and found that the lifetimes could be thousands of time longer than nearest neighbor tunneling times \cite{Stojanovic2008}. 

 As a step towards a complete theory of the metastability of the excited band, we present a detailed characterization of the preparation and relaxation of one of the simplest excited band states, the 1D ``\textit{p}-orbital insulator''. We define this state and describe in detail its preparation for the case of a Gaussian pulse in sections \ref{sec:qual} and \ref{sec:loadrelax}. We compute the initial decay rate $\Gamma$ using Fermi's Golden Rule for a wide range of well depths in section \ref{sec:initstage}. On long time scales our system will evolve to a thermal Bose-Einstein distribution; we compute its parameters in section \ref{sec:thermeq}. In section \ref{sec:longtime}, we study the late stages of relaxation to this equilibrium using a Boltzmann equation formalism. 

Throughout this work we use the recoil energy $E_R=h^2/2m\lambda^2$, where $m$ and $\lambda$ are respectively the atomic mass and wavelength of the laser creating the lattice. Note the optical lattice separation is $\lambda/2$, since the lattice potential is proportional to the intensity of the laser. When we make comparisons to experiment, we use the parameters from \cite{Muller2007} where $E_R=h\times3.273$ kHz. 

\section{\label{sec:qual}Qualitative considerations and main results}

\subsection{Adiabatic loading of the excited band}
\begin{figure}
 \centering
 \includegraphics[width=1.00\columnwidth]{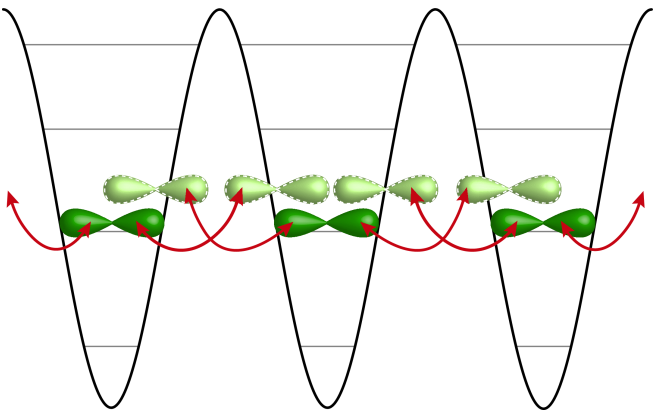}
\caption{(Color online.) A sketch of the $p$-orbital insulator at a density of one particle per site. For the $p$-orbital insulator the hopping energy in the excited band $t_1$ is much smaller than on the on-site interaction $U_{11}$. The density of virtually double occupied sites $\rho\approx 2t_1^2/U_{11}^2$; see Eq.~(\ref{eq:ground_state}). The darker-green orbitals indicate the dominant on-site density, while the lighter-green orbitals indicate virtual hops to the left and right.}
 \label{fig:p_orb_ins}
\end{figure}

A condensate loaded into the ground band $s$ orbitals of an optical lattice at a density of one boson per site will form a Mott insulator for sufficiently deep wells where the on-site interaction is much stronger than the tunneling \cite{Greiner02,Fishers89}. Starting from that ground state, {\em every} atom can be excited to the $p_x$ orbital of the first excited band. In a typical optical lattice setup, the lattice potential in the $y-$ and $z-$ directions is much stronger than in the $x-$ direction; this lifts the degeneracy among the three $p$ orbitals and makes it possible to freeze out excitations to the $p_y$ and $p_z$ orbitals. On time scales short compared to the excited band lifetime $1/\Gamma$, the system is well described by an effective excited band Bose Hubbard Hamiltonian. If the nearest neighbor hopping energy $t_1$ is sufficiently small compared to the on-site interaction $U$, the system will have a ground state with an excitation gap of order $U$. A schematic of the $p$-orbital insulator is shown in Fig. \ref{fig:p_orb_ins}.

To be more explicit, a laser pulse causing a stimulated Raman transition can be used to promote bosons from the ground band to the $p$-orbital insulator \cite{Muller2006}. The optical lattice potential is anharmonic, so the energy difference between the ground and first excited bands $\epsilon_{10}$ is larger than the energy difference between the first and second excited bands $\epsilon_{21}$. The applied laser pulse can be considered adiabatic if its corresponding Rabi frequency is small compared to the anharmonicity $\epsilon_{10}-\epsilon_{21}$. Allowing hops of atoms between the wells of the optical lattice modifies the spectrum of the many-body atomic system by introducing the small hopping energy scale $t_1$. 

A simple system of two atoms on two sites can elucidate many aspects of that modification; we discuss this case in Section \ref{sec:twosite_model}. First, virtual delocalization of the atoms from their respective sites results in the transition frequency shifting from its nominal value by $O(t_1^2/U)$. Second, the evolution of the system under the pulse is characterized by two time scales : the inverse of the on-site interaction $1/U$ and the inverse of the frequency splitting $U/t_1^2$. In the case of a simple two-atom system, the amplitude of the $p$-state formed by a $\pi$-pulse of length $\tau$ initially grows with $\tau$ reaching a local maximum at $\tau\sim\frac{1}{U}\log\frac{U}{t_1}$. The infidelity for a pulse with that optimal pulse length is of order $\frac{t_1^4}{U^4}\log^2\frac{U}{t_1}$. Further increase of the pulse length first increases the infidelity because of the optical nutation between the two-atomic $p$-state and a state in which only one of the atoms is in that state. In the case of two atoms the full fidelity is reached asymptotically for $\pi$-pulse lengths exceeding significantly the $t_1^2/U$ scale. 

In a lattice of many atoms, the $p$-state acquires a finite lifetime $1/\Gamma$. For a typical experimental setup, $\Gamma\ll U$, but $\Gamma\sim t_1^2/U$. The latter relation makes it practically hard to get arbitrarily close to the ground-state of the $p$-state Mott insulator. We argue that the smallest infidelity one may reach before the relaxation of the $p$-state sets in is of the order of $\frac{t_1^4}{U^4}\log^2\frac{U}{t_1}$ (see Section \ref{sec:full_adiabatics}). 

\subsection{Stages of Relaxation}
The $p$-orbital insulator ground state has a fraction of double-occupied sites of $O(t_1/U)^2$. The on-site interaction causes pairs of particles to make transitions out of the $p$-band, releasing large amounts $(\sim\epsilon_{10})$ of energy. Metastability is guaranteed if however the fraction of these doubly occupied sites is sufficiently small. In section \ref{sec:initstage}, we present a detailed calculation for the initial decay of the $p$ orbital insulator. 

We can estimate the initial decay rate $\Gamma$ by applying the Fermi Golden Rule to transitions transferring two atoms from the $p$-band ($n=1$) into states in band $n=2$ and in band $n=0$. The rate $\Gamma$ is proportional to the product of the probability of double occupancy of a state, $\sim (t_1/U)^2$, the square of the inter-band transition matrix element $\sim U^2$, and the combined density of final states. The latter one is $\sim 1/t_2$ and is essentially controlled by the wider band $n=2$. The leading order result for the initial decay rate is
\begin{equation}
\label{eq:gamm_leading_order}
 \Gamma\sim t_1^2/t_2.
\end{equation}
Note in Eq.~(\ref{eq:gamm_leading_order}) that the decay rate $\Gamma$ does not depend on the interaction strength to leading order. For $V>26E_R$, the bandwidth of the second excited band becomes too narrow for this two-body decay process to conserve energy; our analysis predicts a jump discontinuity in the decay rate at the threshold value of well depth. A more complicated three-body process is the dominant initial decay channel for well depths beyond this threshold.

Assuming that the atoms are thermally isolated in the trap, the
very large energy deposited in the system by the initial $\pi$-pulse eventually is
transformed into the thermal energy of the atoms in the trap (see section \ref{sec:thermeq} for details). At equilibrium, the effective temperature $T$ is on the scale of the highly energetic $\epsilon_{10}$. Because of the high thermal energy per atom, the equilibrium density of the system is considerably lower than the initial density due to the spreading of atoms within the magnetic trap. We treat the asymptotic approach to this final equilibrium using a Boltzmann equation formalism in section \ref{sec:longtime}. 

\section{\label{sec:loadrelax}Adiabatic Loading of the excited many-body state}
We want the $\pi$-pulse to excite $N$ bosons from the true ground state to the ground state within the $p_x$ band with high fidelity. In order to elucidate the effect of the length of the pulse we first investigate a model of two sites with two energy levels mixed by tunneling. We then comment on the full $N$-site problem, and show that the condition for adiabaticity remains $U\tau\gg\hbar$, in the sense that the density of excitations out of the $p_x$ band ground state is small in this limit. 

\subsection{Two-site model}
\label{sec:twosite_model}
\begin{figure}
 \centering
 \includegraphics[width=0.80\columnwidth]{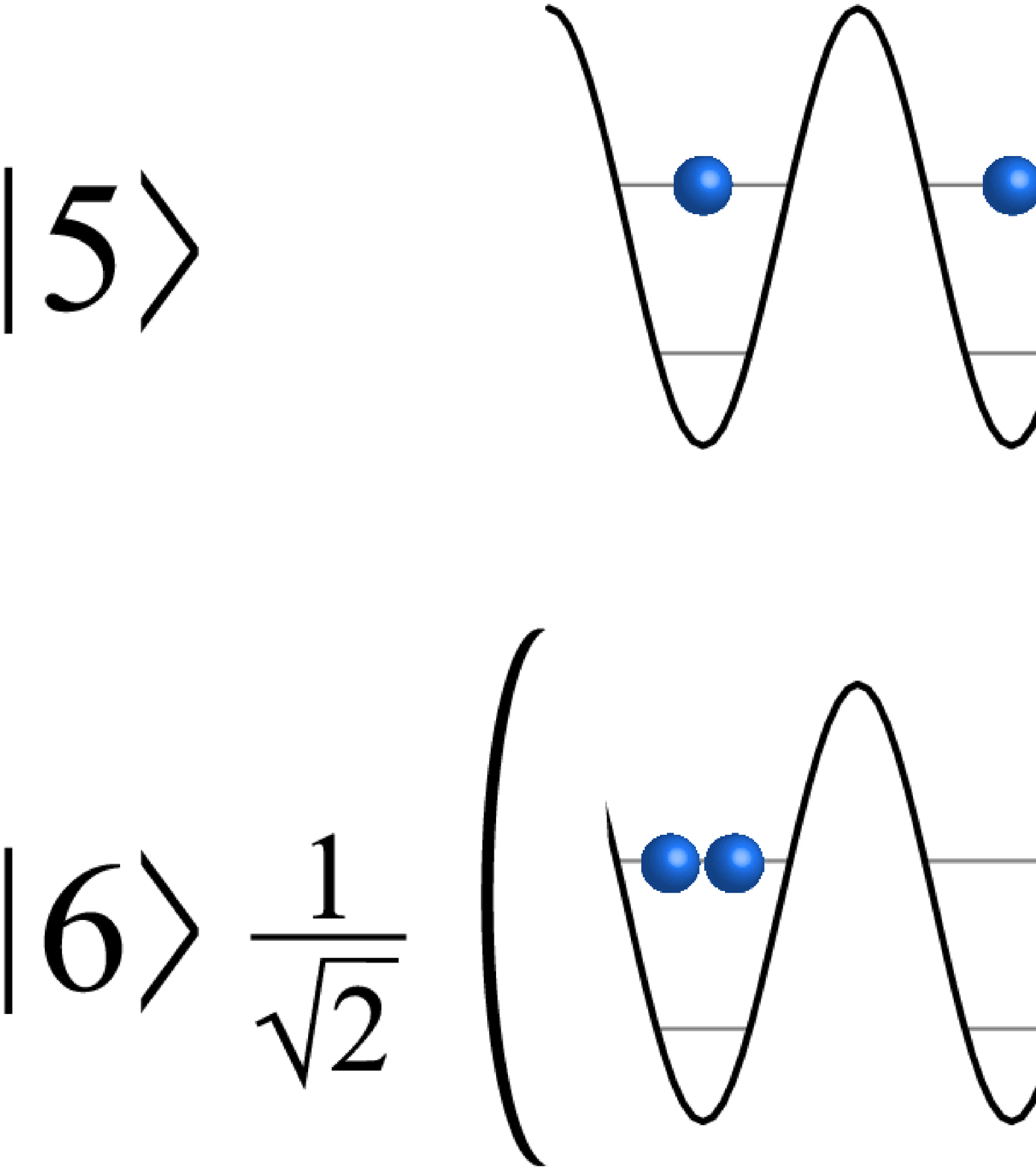}
 \caption{(Color online.) The six parity-symmetric basis states of the two-site model in section \ref{sec:twosite_model}. When the excited band hopping energy $t_1=0$, these states are true eigenstates. For $t_1\neq0$, states $\vert3\rangle$ and $\vert4\rangle$ are mixed, as are states $\vert5\rangle$ and $\vert6\rangle$. The eigenstates are given by Eq.~(\ref{eq:symm_eigenstates}) to leading order in $t_1/U$.}
 \label{fig:basis_states}
\end{figure}

Consider a model of two bosons in a well of two sites. Each site has an $s$ level and a $p$ level, but we only include tunneling in the $p$-band. We denote the band splitting by $\epsilon_{10}$, the excited band hopping energy by $t_{1}$, and the on-site interaction between bosons in bands ${n_1,n_2}$ as $U_{n_1n_2}$. We will describe our model in the second quantized formalism, using the bosonic operators $b_{L0},b_{L1},b_{R0},b_{R1}$ where $\lbrace L,R\rbrace$ refers to the left and right positions and $\lbrace0,1\rbrace$ refers to the band index. Taking the $s$ orbital energy to be zero, the Hamiltonian $H$ is given by the expression:
\begin{align}
\label{eq:Hamiltonian}
H=&\epsilon_{10}b_{L1}^{\dag}b_{L1}+t_{1}b_{L1}^{\dag}b_{R1}+\frac{U_{11}}{2}b_{L1}^{\dag}b_{L1}^{\dag}b_{L1}b_{L1}\nonumber\\
&+\frac{U_{00}}{2}b_{L0}^{\dag}b_{L0}^{\dag}b_{L0}b_{L0}+U_{10}b_{L1}^{\dag}b_{L0}^{\dag}b_{L1}b_{L0}\nonumber\\
&+\left(L\rightarrow R\right).
\end{align}
As our initial state was symmetric with respect to left-right exchange, the relevant Hilbert space consists only of parity-symmetric states. We have sketched these states in Fig. \ref{fig:basis_states}. Neglecting terms of $O(t_1^2/U^2)$, the six parity-symmetric eigenstates of $H$ are 
\begin{align}
\left|1\right\rangle = & b_{L0}^{\dag}b_{R0}^{\dag}\left|0\right\rangle, \nonumber\\
\left|2\right\rangle = & \frac{1}{\sqrt{2}}\left(\frac{1}{\sqrt{2!}}b_{L0}^{\dag}b_{L0}^{\dag}+\frac{1}{\sqrt{2!}}b_{R0}^{\dag}b_{R0}^{\dag}\right)\left|0\right\rangle, \nonumber\\
\left|3\right\rangle = & \frac{1}{\sqrt{2}}\left(b_{L1}^{\dag}b_{R0}^{\dag}-\frac{t_{1}}{U_{10}}b_{L1}^{\dag}b_{L0}^{\dag}+L\rightarrow R\right)\left|0\right\rangle, \nonumber\\
\left|4\right\rangle = & \frac{1}{\sqrt{2}}\left(\frac{t_{1}}{U_{10}}b_{L1}^{\dag}b_{R0}^{\dag}+b_{L1}^{\dag}b_{L0}^{\dag}+L\rightarrow R\right)\left|0\right\rangle, \nonumber\\
\left|5\right\rangle = & b_{L1}^{\dag}b_{R1}^{\dag}-\frac{2t_{1}}{U_{11}}\frac{1}{\sqrt{2}}\left(\frac{1}{\sqrt{2!}}b_{L1}^{\dag}b_{L1}^{\dag}+\frac{1}{\sqrt{2!}}b_{R1}^{\dag}b_{R1}^{\dag}\right)\left|0\right\rangle, \nonumber\\
\left|6\right\rangle = & \frac{2t_{1}}{U_{11}}b_{L1}^{\dag}b_{R1}^{\dag}+\frac{1}{\sqrt{2}}\left(\frac{1}{\sqrt{2!}}b_{L1}^{\dag}b_{L1}^{\dag}+\frac{1}{\sqrt{2!}}b_{R1}^{\dag}b_{R1}^{\dag}\right)\left|0\right\rangle.
\label{eq:symm_eigenstates}
\end{align}
The corresponding eigenvalues for these states to the same order in $t_1/U$ are
\begin{align}
\tilde{H}_{11}=&0, &  \tilde{H}_{22}=&U_{00},\nonumber\\
\tilde{H}_{33}=&\epsilon_{1}-\frac{t_{1}^{2}}{U_{10}}, &  \tilde{H}_{44}=&\epsilon_{1}+U_{10}+\frac{t_{1}^{2}}{U_{10}},\nonumber\\
\tilde{H}_{55}=&2\epsilon_{1}-\frac{4t_{1}^{2}}{U_{11}}, &  \tilde{H}_{66}=&2\epsilon_{1}+U_{11}+\frac{4t_{1}^{2}}{U_{11}}.\end{align}
Within our model, we want to prepare the system in state $\left|1\right\rangle$ and, by applying a $\pi$-pulse, adiabatically evolve it into state $\left|5\right\rangle$. This process is the two-site equivalent of promoting a ground band Mott Insulator into the \textit{p}-orbital insulator. We consider a general pulse term of the form:
\begin{equation}
\label{eq:hpulse}
H_{\textrm{pulse}}=\Delta(t)\ e^{-i\omega t}\left(b_{L1}^{\dag}b_{L0}+b_{R1}^{\dag}b_{R0}\right)+\textrm{h.c.},
\end{equation}
with the Gaussian envelope $\Delta(t)=\frac{\alpha}{\tau}\exp(-t^2/\tau^2)$. We will choose the dimensionless pulse strength $\alpha$ to optimize a single $\pi$-pulse . As the excitation can be considered a standard two-photon process, the carrier frequency of the pulse $\omega$ is chosen to be half the energy difference of states $\left|1\right\rangle$ and $\left|5\right\rangle$:
\begin{subequations}
\begin{align}
\omega&=\frac{1}{2}\left(\tilde{H}_{55}-\tilde{H}_{11}\right),\\
&=\epsilon_{1}-\frac{2t_{1}^{2}}{U_{11}}.
\end{align}
\end{subequations}
To simplify notation in the following discussion, we define the dimensionless time $u=t/\tau$, the dimensionless on-site interaction in the $p$-band $\tilde{U}=U_{11}\tau$ and the dimensionless detuning $\kappa$ between the energy of intermediate state $\vert 3 \rangle$ and the drive frequency:
\begin{subequations}
\label{eq:dimensionless_variables}
\begin{align}
\kappa&=(\tilde{H}_{33}-\omega)\tau,\\
&=\frac{t_1^2}{U_{11}}\tau\left(2-\frac{U_{11}}{U_{10}}\right).
\end{align}
\end{subequations}
Experimentally, $\kappa$ and $\tilde{U}$ can be controlled nearly independently since $\kappa$ depends much more sensitively on well depth than $\tilde{U}$. In our calculation, we will treat these two parameters as independent. A conversion table between $\lbrace\kappa,\tilde{U}\rbrace$ and the more experimentally relevant variables of well depth $\tilde{V}$ and pulse time $\tau$ are given in Table \ref{tab:dictionary}.

\begin{table}
\centering
\begin{tabular}{|c|c|c|c|c|c|c|c|c|}
\hline 
$\kappa$&$\tilde{U}$&$\tilde{V} (E_R)$&$\tau$(ms)&
&$\kappa$&$\tilde{U}$&$\tilde{V}(E_R)$&$\tau$(ms)\tabularnewline\hline
\hline 
1.0E-7&0.1&54.33&0.0342&&0.01&0.1&13.41&0.0566\tabularnewline\hline 
&&&&&0.01&1&20.25&0.4750\tabularnewline\hline 
1.0E-6&0.1&44.72&0.0363&&0.01&10&27.69&4.236\tabularnewline\hline 
1.0E-6&1.0&54.33&0.3418&&0.01&100&35.84&38.87\tabularnewline\hline 
&&&&&&&&\tabularnewline\hline 
1.0E-5&0.1&35.84&0.0389&&0.1&0.1&6.65&0.0764\tabularnewline\hline 
1.0E-5&1&44.72&0.3626&&0.1&1&13.41&0.5655\tabularnewline\hline 
1.0E-5&10&54.33&3.418&&0.1&10&20.25&4.750\tabularnewline\hline 
&&&&&0.1&100&27.69&42.36\tabularnewline\hline 
1.0E-4&0.1&27.69&0.0423&&&&&\tabularnewline\hline 
1.0E-4&1&35.84&0.3887&&1&1&6.65&0.7644\tabularnewline\hline 
1.0E-4&10&44.72&3.626&&1&10&13.41&5.655\tabularnewline\hline 
&&&&&1&100&20.25&47.50\tabularnewline\hline 
0.001&0.1&20.25&0.0047&&&&&\tabularnewline\hline 
0.001&1&27.69&0.4236&&10&10&6.65&7.644\tabularnewline\hline
0.001&10&35.83&3.887&&10&100&13.41&56.55\tabularnewline\hline
0.001&100&44.71&36.26&&&&&\tabularnewline\hline
\end{tabular}
\caption{A conversion table between the dimensionless parameters $\tilde{U}=U_{11}\tau$ and $\kappa$ defined in Eq.~(\ref{eq:dimensionless_variables}) and the experimentally relevant values of well depth $\tilde{V}$ and pulse time $\tau$. For fixed pulse time, $\kappa$ is exponentially sensitive to changes in well depth whereas $\tilde{U}$ is nearly constant over a wide range of well depth energies; this difference in sensitivity allows an experimentalist to tune the two parameters nearly independently. }
\label{tab:dictionary}
\end{table}

To discuss the state's time evolution, we will write a Schr\"odinger equation for the amplitudes of each of the six parity-symmetric eigenvectors: 
\begin{align}
\left|\psi(t)\right\rangle=a_{1}\left|1\right\rangle +a_{2}\left|2\right\rangle+e^{i\omega t}a_{3}\left|3\right\rangle+e^{i\omega t}a_{4}\left|4\right\rangle\nonumber
\\+e^{i2\omega t}a_{5}\left|5\right\rangle +e^{i2\omega t}a_{6}\left|6\right\rangle.
\end{align}
The Hilbert space can be considered as two independent subspaces $\lbrace \left\vert1\right\rangle,\left\vert3\right\rangle,\left\vert5\right\rangle \rbrace$ and $\lbrace \left\vert2\right\rangle,\left\vert4\right\rangle,\left\vert6\right\rangle \rbrace$ weakly coupled by the pulse. The amplitudes of the second subspace are much smaller because their corresponding energies are shifted by order the Mott gap compared to the energies of the first subspace. The strategy for solution in the $\kappa \ll 1$ limit consists of restricting the evolution to the $\lbrace \left\vert1\right\rangle,\left\vert3\right\rangle,\left\vert5\right\rangle \rbrace$ subspace, and then using this solution as a drive term to find the evolution of the much smaller amplitudes in the $\lbrace \left\vert2\right\rangle,\left\vert4\right\rangle,\left\vert6\right\rangle \rbrace$ subspace. The details are located in Appendix \ref{sec:appendix_pulse}.

The important parameters to determine from our analysis are the dimensionless strength of the optimal pulse $\alpha$, the final amplitude of the \textit{p}-orbital insulator $\vert a_{5}(\infty)\vert^2$ subject to this optimal pulse, and the amplitudes $\lbrace a_2,a_4,a_6\rbrace$. In the $\kappa \ll 1$ and $\tilde{U}\gg1$ limit, we can expand these parameters in the small parameters $\kappa$ and $1/\tilde{U}$ :
\begin{subequations}
\label{eq:a5_val}
\begin{align}
&\left|a_n(\infty)\right|^2=\frac{\kappa}{\tilde{U}}\left|a_n^{(1)}\right|^2+O\left(\frac{\kappa^2}{\tilde{U}^2}\right) \text{ for }n=2,4,6\ \ ,\\
&\left|a_5(\infty)\right|^2=1-\frac{\kappa}{\tilde{U}}\left(\left|a_2^{(1)}\right|^2+\left|a_4^{(1)}\right|^2+\left|a_6^{(1)}\right|^2\right)\nonumber\\
&\ \ \ \ \ \ \ \ \ \ -(0.265\ldots)\kappa^2+(0.0045\ldots)\kappa^4+O\left(\kappa^6,\frac{\kappa^2}{\tilde{U}^2}\right),\\
&\alpha=\frac{\sqrt{\pi}}{2}-(0.0425\ldots)\kappa^2+O\left(\kappa^4\right).
\end{align}
\end{subequations}
The coefficients $\lbrace a_{2}^{(1)},a_{4}^{(1)},a_{6}^{(1)}\rbrace$ are calculated in Appendix \ref{sec:scratchwork_subspace_two}:
\begin{subequations}
\begin{align}
 \left|a_2^{(1)}\right|^2&=\frac{\pi r^2}{8\left(2-r\right)}\left|\int_{-\infty}^{\infty}due^{-u^2+iU_2u}\cos\left(\frac{\pi \textrm{erf }u}{2}\right)\right|^2,\\
\left|a_4^{(1)}\right|^2&=\frac{2\pi}{2-r}\left|\int_{-\infty}^{\infty}du\left\{\frac{r}{2}\cos^2\left[\frac{\pi\left(1+\textrm{erf }u\right)}{4}\right]\right.\right.\nonumber\\
&\left.\left.+\left(1-\frac{r}{2}\right)\sin^2\left[\frac{\pi\left(1+\textrm{erf }u\right)}{4}\right]\right\}e^{-u^2+iU_4u}\right|^2,\\
\left|a_6^{(1)}\right|^2&=\frac{\pi \left(1-\frac{r}{4}\right)^2}{\left(1-\frac{r}{2}\right)}\left|\int_{-\infty}^{\infty}due^{-u^2+iU_6u}\cos\left(\frac{\pi \textrm{erf }u}{2}\right)\right|^2,
 \end{align}
\end{subequations}
with $r=U_{11}/U_{10}$ and the dimensionless interaction energies $\lbrace U_2,U_4,U_6 \rbrace$ defined by
\begin{subequations}
\begin{align}
&U_{2}=\frac{U_{00}}{U_{11}}\tilde{U},\\
&U_{4}=r^{-1}\tilde{U}+\kappa\left(\frac{2+r}{2-r}\right),\\
&U_6=\tilde{U}+\frac{8\kappa}{2-r}.
\end{align}
\end{subequations}
Among the exponentially suppressed amplitudes $\lbrace a_2,a_4,a_6\rbrace$, by far the largest is $a_6$. As shown in Appendix \ref{sec:scratchwork_subspace_two}, the dominance of $a_6$ is a consequence of $U_{11}$ being smaller than either $U_{10}$ or $U_{00}$. 

An important observation about Eq.~(\ref{eq:a5_val}) is that perfect fidelity is impossible for a finite $\kappa$ even with $\tilde{U}\rightarrow\infty$. This occurs because the spectral width of the pulse is sufficient to populate the intermediate state. The Hamiltonian $H+H_{\textrm{pulse}}$ in the limit $\tilde{U}\rightarrow\infty$ is similar to that of a spin-$1$ system in a magnetic field, where a small quadratic Zeeman shift leads to nonequidistant energy levels (see eqs.\ref{eq:Hamiltonian} and \ref{eq:hpulse}). In principle, if one would increase the pulse length and went to sufficiently large values of $\kappa$, one could again approach high fidelities with a single frequency. This limit is infeasible, however, due to the finite lifetime of the excited band. 

\begin{figure}
 \centering
 \includegraphics[width=1.00\columnwidth]{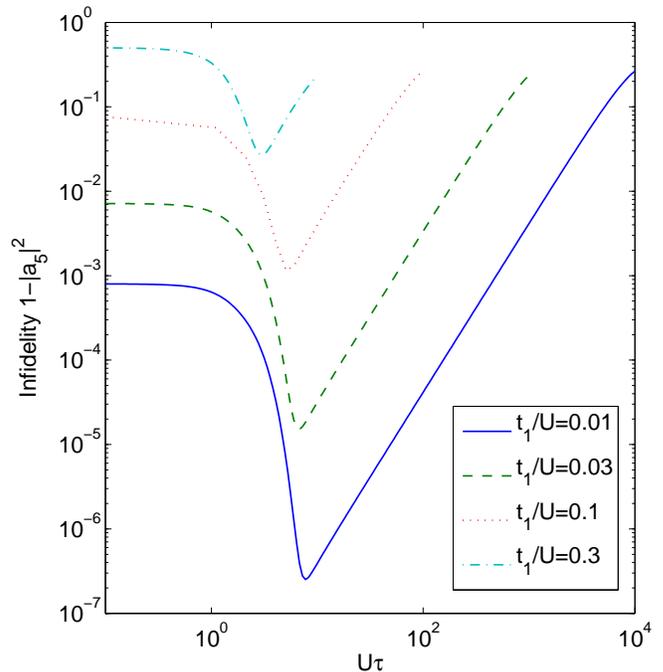}
\caption{Probability of not finishing in the two-atom $p$-state after application of a $\pi$-pulse ($1-|a_5|^2$) plotted versus $U\tau$ at fixed $t_1/U$; see Eq.~(\ref{eq:a5_val}). The curve minimum (corresponding to maximum fidelity) is determined by two competing effects. Increasing $U\tau$ exponentially suppresses $\left|a_2\right|^2$, $\left|a_4\right|^2$ and $\left|a_6\right|^2$, but also increases $\left|a_1\right|^2$ and $\left|a_3\right|^2$. The minimum occurs at $U\tau\sim\log(U/t_1)$ and has value $O\left(\frac{t_1^4}{U^4}\log^2\frac{U}{t_1}\right)$; see Eqs.~(\ref{eq:tau_opt}) and (\ref{eq:a5_minval}). }
\label{fig:a5diff_fixed_toveru}
\end{figure}

As shown in Fig. \ref{fig:a5diff_fixed_toveru}, for a given well depth there is an optimal pulse time $\tau$ which balances the decreasing infidelity at larger $\tilde{U}$ against the increasing infidelity at larger $\kappa$. We can approximate the optimal pulse by using the following rough form for $1-\left|a_5\right|^2$:
\begin{equation}
\label{eq:a5_diff_rough}
 1-\left|a_5\right|^2\approx a e^{-2U\tau}+b \frac{t_1^4}{U^2}\tau^2,
\end{equation}
with $a,b\sim O(1)$. The first term comes from the suppression of $|a_6|^2$, and the second from the growth of $|a_1|^2$ and $a_3^2$. The asymptotic expansion of the optimal pulse time $\tau_{\textrm{opt}}$ in Eq.~(\ref{eq:a5_diff_rough}) for small $t_1/U$ is
\begin{equation}
\label{eq:tau_opt}
 \tau_{\textrm{opt}}=\frac{2}{U}\log\frac{U}{t_1}+O\left(\frac{1}{U}\log\log\frac{U}{t_1}\right).
\end{equation}
The expression for the resulting infidelity is
\begin{equation}
\label{eq:a5_minval}
 \left|a_5(\tau_{\textrm{opt}})\right|^2\approx \frac{4bt_1^4}{U^4}\log^2\frac{U}{t_1}.
\end{equation}
 
A two-tone pulse could make $\left|a_5(\infty)\right|^2$ arbitrarily close to unity through the application of two $\pi$-pulses, but for small $\kappa$ the fidelity from a single-tone pulse is already better than the $80\%$ transfer efficiency observed in experiment \cite{Muller2007}. The observed infidelities are more than two orders of magnitude larger than those calculated here. This striking difference could potentially come from the experimental pulse shape not being close to Gaussian or the internal structure of the Rb atom interfering with pulse fidelity. 

\subsection{Full N-site Adiabatics}
\label{sec:full_adiabatics}
\begin{figure}
 \centering
 \includegraphics[width=1.00\columnwidth]{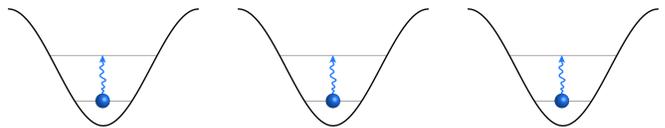}
 \caption{(Color online.) We generalize to the case of a $\pi$-pulse on $N$ sites by first solving for the evolution of the system in the absence of tunneling, and then treat tunneling as a perturbation in the interaction picture.}
 \label{fig:nsite_adiabatics}
\end{figure}

We now consider the problem of loading the excited band in an system of $N$ sites, modeling each site as a two-level system with energy splitting $\epsilon_{10}$ and with tunneling of energy $t_1$ permitted between the excited levels of neighboring sites. Additionally, we make the important simplification of assuming all the Hubbard $U$ parameters are equal. Initially, the system is prepared in its true ground state: one particle per site, each site in the lower energy level. An excitation pulse $\Delta(t)e^{i\omega t}$ promotes bosons from the ground band to the first excited band (see Fig. \ref{fig:nsite_adiabatics}).

 We are interested in pulses sufficiently long that $\tilde{U}\gg1$, so that the excitation density above the ground state is small. We refer to such pulses as ``adiabatic'', although this is not the conventional meaning of a slow external process which transfers no heat to the system \cite{Landau58}. The three important pulse parameters are its carrier frequency $\omega$, its dimensionless strength $\alpha$, and its duration $\tau$. We choose the carrier frequency to maximize the occupation of the $p$-orbital insulator and adjust the pulse strength to produce a $\pi-$pulse, leaving the pulse duration as the most important adjustable parameter. 

\begin{figure}
 \centering
 \includegraphics[width=1.00\columnwidth]{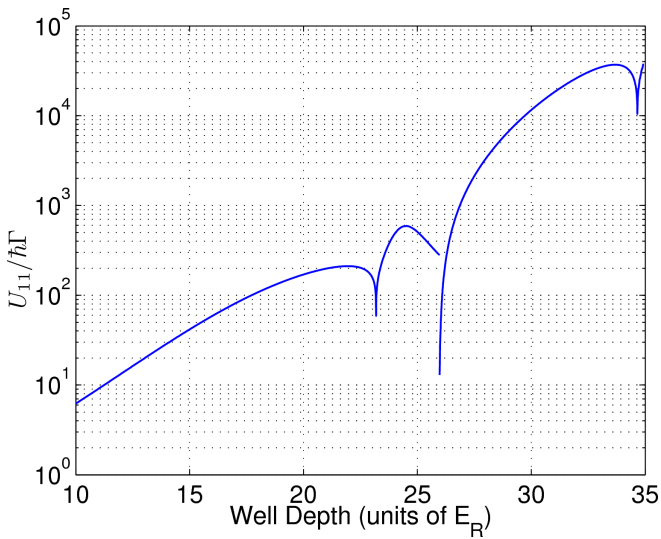}
\caption{Ratio of on-site interaction energy to decay rate versus well depth. This ratio must be large to justify a key assumption of this work: pulse times can both be sufficiently long to treat the system in the the adiabatic regime ($\tau\gg1/U$) and also respect the finite lifetime of the excited band ($\tau<1/\Gamma$).}
\label{fig:interaction_decay_ratio}
\end{figure}

We require $U_{11}/\Gamma\gg1$ so that the pulse time $\tau$ is both much longer than $1/U$ from adiabatics considerations and also shorter than $1/\Gamma$ from finite lifetime considerations. This condition is satisfied for a wide range of well depths (see Fig. \ref{fig:interaction_decay_ratio}).

The Hamiltonian $H$ for this system in the frame rotating at the carrier frequency of the pulse $\omega$ is
\begin{subequations}
\begin{align}
 H=&H_0(t)+H_1,\\
H_0=&\sum_j \Delta(t)\left(b_{1j}^{\dag}b_{0j}+b_{0j}^{\dag}b_{1j}\right),\\
H_1=&\left(\epsilon_{10}-\omega\right)\sum_jb_{1j}^{\dag}b_{1j}+t_1\sum_{\langle ij\rangle}b_{1i}^{\dag}b_{1j}
\\&+\frac{U}{2}\sum_j\sum_{n=0,1}b_{nj}^{\dag}\left(\sum_{n'=0,1}b_{n'j}^{\dag}b_{n'j}\right)b_{nj}.\nonumber
\end{align}
\end{subequations}
The initial state $\left|\psi_0\right\rangle$ is given in second-quantized notation in the soluble limit $t_1=0$ by:
\begin{equation}
\left|\psi_0\right\rangle=\prod_jb_{0j}^{\dag}\left\vert0\right\rangle.
\end{equation}
To find the evolution of the system as a perturbation expansion in the small parameter $t_1/U$, we transform to the interaction picture:
\begin{subequations}
\begin{align}
\label{eq:int_schrod}
i\frac{d}{dt}\left|\Psi_I\right\rangle=&H_I\left|\Psi_I\right\rangle,\\
H_I(t)=&e^{i\int_{-\infty}^{t}dt'H_{0}(t')}H_1e^{-i\int_{-\infty}^{t}dt'H_{0}(t')}.
\end{align}
\end{subequations}
Since we are treating all the Hubbard $U$ terms as equal, the on-site interaction commutes with $H_0$. In the transformed basis, $H_I$ becomes 
\begin{subequations}
\begin{align}
\label{eqnarray:Hamiltonian_interaction_picture}
H_I(t)=&\frac{U}{2}\sum_j\sum_{nn'}b_{nj}^{\dag}b_{n'j}^{\dag}b_{n'j}b_{nj}+t_1\sum_{\langle jj'\rangle}B_j^{\dag}(t)B_{j'}(t)\nonumber\\
&+\left(\epsilon_{10}-\omega\right)\sum_jB_j^{\dag}(t)B_j(t),\\
B_j^{\dag}(t)=&\cos\theta b_{1j}^{\dag}+i\sin\theta b_{0j}^{\dag},
\end{align}
\end{subequations}
with the parameter $\theta$ given by
\begin{equation}
\theta=\int_{-\infty}^tdt'\Delta(t').
\end{equation}
By changing to the interaction picture, we have transformed the time dependence of the pulse into time dependence of the tunneling and detuning $\epsilon_{10}-\omega$. We choose the duration of the pulse $\tau$ to correspond to a $\pi$-pulse when the frequency is resonant for a single site $\omega=\epsilon_{10}$. There are two parameters we would like to determine: the probability $P_g$ that the final state after application of the pulse is the $p$-orbital insulator and the ``density of excitations'' $\rho_{\textrm{exc}}$ of the final state. 

We assume for simplicity that every particle is promoted to the $p$-band. Since an excitation above the $p$-band ground state costs approximately $U$ in energy for $t_1/U\ll1$, we can define $\rho_{\textrm{exc}}$ as \cite{Gritsev2008}
\begin{equation}
\label{eq:rho_exc_def}
\rho_{\textrm{exc}}=\frac{\left\langle\Psi_I(\infty)\right|H_I(\infty)\left|\Psi_I(\infty)\right\rangle-\left\langle\Psi_g\right|H_I(\infty)\left|\Psi_g\right\rangle}{NU}.
\end{equation}
We note that Eq.~(\ref{eq:rho_exc_def}) does not simply count the density of doubly excited states, since the $p$-orbital insulator has $O(t_1^2/U^2)$ doubly excited states. Rather Eq.~(\ref{eq:rho_exc_def}) measures the relative potential energy of the final state in units of the on-site interaction $U$.

We cannot simply apply the perturbation methods of the previous section to the $N$-site model because the expansion parameter changes from $t_1/U$ to $t_1\sqrt{N}/U$, which leads to a series that is not obviously convergent in the thermodynamic limit. Instead, we approach the problem using only intensive quantities at every stage. Consider turning on the tunneling between nearest neighbor sites one at a time. The Hamiltonian $H_n$ with hopping allowed only between the first $n$ nearest neighbor pairs is
\begin{align}
\label{eq:hamiltonian_n}
H_n(t)=&t_1\sum_{j=1}^n\left(B_{j+1}^{\dag}B_{j}+\textrm{h.c.}\right)+\left(\epsilon_{10}-\omega\right)\sum_jB_j^{\dag}B_j\nonumber\\
&+\frac{U}{2}\sum_j\sum_{nn'}b_{nj}^{\dag}b_{n'j}^{\dag}b_{n'j}b_{nj}.
\end{align}
To compute the probability $P_g$ that the system ends in the ground state and the excitation density $\rho_{\textrm{exc}}$, we derive recursion relations describing how these quantities change when tunneling is activated between one additional nearest neighbor pair. Solving these recursion relations, we take $n\rightarrow N$ to derive results that remain valid in the thermodynamic limit where $t_1\sqrt{N}/U$ is a large parameter. This removes many of the difficulties present in Rayleigh-Schr\"odinger perturbation theory. 

\begin{figure}
 \centering
 \includegraphics[width=1.00 \columnwidth]{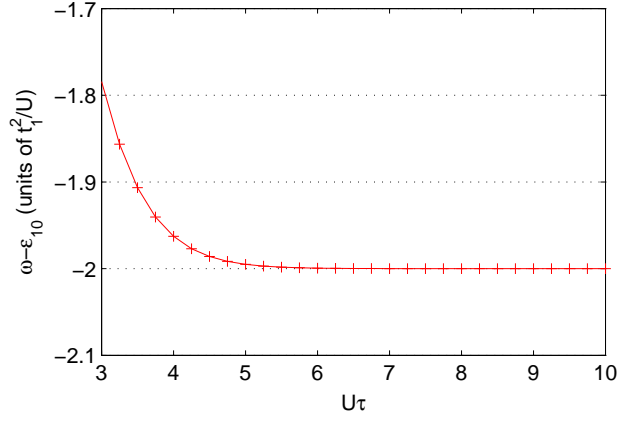}
\caption{The optimal pulse carrier frequency $\omega-\epsilon_{10}$ in units of $t_1^2/U$ for the specific case of a Gaussian pulse; see Eq.~(\ref{eq:pulse_freq_opt}). The optimal frequency maximizes the occupation of the $p$-orbital insulator after completion of the $\pi$-pulse. As expected, for $U\tau\gg1$ we recover the two-site result that $\omega\approx\epsilon_{10}-\frac{2t_1^2}{U}$. }
\label{fig:pulse_error}
\end{figure}

We begin our discussion with $\left\vert\Psi_g^n\right\rangle$, the ground state wave function of $H_n(\infty)$:
\begin{align}
\label{eq:ground_state}
 \left|\Psi_g^n\right\rangle=&\prod_{j=1}^Nb_{0j}^{\dag}\left\vert0\right\rangle-\frac{t_1}{U}\sum_{j=1}^n\left(b_{0j+1}^{\dag}b_{0j}+\textrm{h.c.}\right)\prod_{j=1}^Nb_{0j}^{\dag}\left\vert0\right\rangle\nonumber\\&+O\left(\frac{t_1^2}{U^2}\right).
\end{align}
We see from Eq.~(\ref{eq:ground_state}) that the virtually double occupied fraction is $O(t_1^2/U^2)$ as illustrated in Fig. \ref{fig:p_orb_ins}. As discussed in Appendix \ref{sec:diff_eqns_appendix}, we can determine the ground state wave function by enforcing the adiabatic pulse condition:
\begin{equation}
\lim_{\tilde{U}\rightarrow\infty}\left\vert\Psi_I^n\right\rangle=\left|\Psi_g^n\right\rangle,
\end{equation}
where $\left\vert\Psi_I^n\right\rangle$ is the final many-body wave function in the interaction picture:
\begin{equation}
 \label{eq:time_development}
\left\vert\Psi_I^n\right\rangle=\exp\left(-iT\int_{-\infty}^{t} dt'H_n(t')\right)\left|\psi_0\right\rangle,
\end{equation}
and $T$ is the Dyson time-ordering operator. Expanding Eq.~(\ref{eq:time_development}) to leading order in $H_n$ gives 
\begin{align}
\label{eq:psi_i_n}
\left|\Psi_I^{n}\right\rangle&=A_{n}(\infty)\left|\Psi_g^{n}\right\rangle\nonumber\\
&+\frac{t_1}{U\sqrt{2}}\delta B^{(1)}(\infty)\sum_{j=1}^{n}\left(b_{1j+1}^{\dag}+b_{1j}\right)b_{0j}\left|\Psi_g^{n}\right\rangle\nonumber\\
&+\frac{t_1}{2U}\delta C^{(1)}(\infty)\sum_{j=1}^{n}\left(b_{0j+1}^{\dag}+b_{0j}^{\dag}\right)b_{0j}\left|\Psi_g^{n}\right\rangle\nonumber\\
&+\frac{t_1^2\tau}{U}D^{(1)}(\infty)\sum_{j=1}^{n}b_{1j}^{\dag}b_{0j}\left|\Psi_g^{n}\right\rangle+O\left(\frac{nt_1^2}{U^2},\frac{nt_1^4\tau^2}{U^2}\right).
\end{align}
As shown in Appendix \ref{sec:diff_eqns_appendix}, the frequency $\omega$ which maximizes the occupation of the $p$-orbital insulator is given by:
\begin{subequations}
\label{eq:pulse_freq_opt}
\begin{align}
\omega\approx&\epsilon_{10}+\frac{t_1^2}{U}\left(-2+\frac{\textrm{Re}\left[I_2\right]}{I_1}\right),\\
I_1=&\int_{-\infty}^{\infty}du\sin\theta(u)\cos\theta(u),\\
I_2=&\int_{-\infty}^{\infty}du\int_{-\infty}^{u}du'e^{i\tilde{U}(u'-u)}\dot{\theta}(u')\nonumber\times\\&\left[2\cos^2\theta (u)\cos 2\theta(u')+\sin2\theta(u)\sin2\theta(u')\right].
\end{align}
\end{subequations}
We have plotted $\omega-\epsilon_{10}$ in units of $t_1^2/U$ in Fig. \ref{fig:pulse_error}. In the limit of large $\tilde{U}$, $\omega\rightarrow\epsilon_{10}-2t_1^2/U$, just as in the two-site case considered in section \ref{sec:twosite_model}. For this particular frequency,
\begin{subequations}
 \begin{align}
\left|\delta B^{(1)}(\infty)\right|^2=&2\left|\int_{-\infty}^{\infty}du\int_{-\infty}^udu'e^{-iUu}\dot{\theta}\cos2\theta e^{iUu'}\right|^2,\\
\left|\delta C^{(1)}(\infty)\right|^2=&4\left|\int_{-\infty}^{\infty}du\int_{-\infty}^udu'e^{-iUu}\dot{\theta}\sin2\theta e^{iUu'}\right|^2,\\
\left|D^{(1)}(\infty)\right|^2=&\left|\int_{-\infty}^{\infty}du\int_{-\infty}^{u}du'\sin\left[\tilde{U}(u'-u)\right]\dot{\theta}(u')\nonumber\times\right.\\
&\bigg\{\cos2\theta(u')+\cos\left[2\theta(u)-2\theta(u')\right]\bigg\}\bigg|^2.
 \end{align}
\end{subequations}
As expected, $\left|\delta B^{(1)}(\infty)\right|^2$, $\left|\delta C^{(1)}(\infty)\right|^2$ and $\left|\delta D^{(1)}(\infty)\right|^2$ are all exponentially suppressed for large $\tilde{U}$. From Appendix \ref{sec:diff_eqns_appendix}, the recursion relations for $P_g$ and $\rho_{\textrm{exc}}$ are
\begin{subequations}
\label{eq:pgn_rhoexcn}
\begin{align}
\log\frac{P_g^{n+1}}{P_g^n}=&-\frac{t_1^2}{U^2}\left(\left\vert \delta B^{(1)}(\infty)\right\vert^2+\left\vert \delta C^{(1)}(\infty)\right\vert^2\right)\nonumber\\
&-\frac{2t_1^4\tau^2}{U^2}\left|D^{(1)}(\infty)\right|^2+O\left(\frac{t_1^4}{U^4},\frac{t_1^6\tau^2}{U^4}\right),\\
\rho_{\textrm{exc}}(n+1)=&\rho_{\textrm{exc}}(n)+\frac{t_1^2}{NU^2}
\left\vert \delta B^{(1)}(\infty)\right\vert^2\nonumber\\
&+\frac{t_1^2}{NU^2}\left\vert \delta C^{(1)}(\infty)\right\vert^2+O\left(\frac{t_1^4}{NU^4}\right).
\end{align}
\end{subequations}
Solving the recursion relations Eq.~(\ref{eq:pgn_rhoexcn}) for $n\rightarrow N$ gives 
\begin{subequations}
\label{eqnarray:excitation_density_form} 
\begin{align}
\rho_{\textrm{exc}}(N)=&\frac{t_1^2}{U^2}\left(\left|\delta B^{(1)}(\infty)\right|^2+\left|\delta C^{(1)}(\infty)\right|^2\right)\nonumber\\
&+O\left(\frac{t_1^4}{U^4}\right),\\
P_g(N)=&\exp\left[-N\frac{t_1^2}{U^2}\left(\left|\delta B^{(1)}(\infty)\right|^2+\left|\delta C^{(1)}(\infty)\right|^2\right)\right.\nonumber\\
&\left.-\frac{2Nt_1^4\tau^2}{U^2}\left|D^{(1)}(\infty)\right|^2+O\left(\frac{Nt_1^2}{U^2},\frac{Nt_1^4\tau^2}{U^2}\right)\right].
\end{align}
\end{subequations}
The exponential form of the ground state probability clearly is consistent with the linked cluster picture that an extensive perturbation theory exponentiates. For the specific case of a Gaussian envelope, $\Delta(t)=\frac{\sqrt{\pi}}{2\tau}\exp(-t/\tau)^2$, the formulas for $\left|\delta B^{(1)}(\infty)\right|^2$, $\left|\delta C^{(1)}(\infty)\right|^2$ and $\left|D^{(1)}(\infty)\right|^2$ are :
\begin{subequations}
\label{eqnarray:excitation_density}
\begin{align}
\left|\delta B^{(1)}(\infty)\right|^2&=\frac{\pi}{2}\left\vert\int_{-\infty}^{\infty}due^{i\tilde{U}u-u^2}\sin\left(\frac{\pi\textrm{erf }u}{2}\right)\right\vert^2,\\
\left|\delta C^{(1)}(\infty)\right|^2&=\pi\left\vert\int_{-\infty}^{\infty}du e^{i\tilde{U}u-u^2}\cos\left(\frac{\pi\textrm{erf }u}{2}\right)\right\vert^2,\\
\left|D^{(1)}(\infty)\right|^2&=\frac{\pi}{2}\left|\int_{-\infty}^{\infty}du\int_{-\infty}^{u}du'e^{-u'^2}\sin\left[\tilde{U}(u'-u)\right]\right.\nonumber\\
&\left.\times\left\{2\cos^2\left[\frac{\pi}{4}\left(1+\textrm{erf }u\right)\right]\sin\frac{\pi\textrm{erf }u'}{2}\right.\right.\nonumber\\
&\left.\left.-\cos\frac{\pi \textrm{erf }u}{2}\cos\frac{\pi \textrm{erf }u'}{2}\right\}\right|^2.
\end{align}
\end{subequations}
We have plotted the sum $\left|\delta B^{(1)}(\infty)\right|^2+\left|\delta C^{(1)}(\infty)\right|^2$ in Fig. \ref{fig:excitation_density}. As shown in Appendix \ref{sec:stat_phase}, the excitation density $\rho_{\textrm{exc}}$ can be estimated for $\tilde{U}\gg1$ using the stationary phase approximation:
\begin{align}
\label{eq:stat_phase_estimate}
\delta B^{(1)}(\infty)\approx\frac{\pi}{4}\left|\frac{Y_{\textrm{stat}}(\tilde{U})-Y_{\textrm{stat}}(-\tilde{U})^*}{2}\right|^2,
\\ \delta C^{(1)}(\infty)\approx\frac{\pi}{2}\left|\frac{Y_{\textrm{stat}}(\tilde{U})+Y_{\textrm{stat}}(-\tilde{U})^*}{2}\right|^2,
\end{align}
with $Y_{\textrm{stat}}(k)$ given by:
\begin{align}
Y_{\textrm{stat}}(\tilde{U})=\frac{\sqrt{\tilde{U}}}{(\log\frac{\tilde{U}}{\sqrt{\pi}})^{1/4}}\exp\left[-\vert \tilde{U}\vert\sqrt{\log\frac{\tilde{U}}{\sqrt{\pi}}}\right.\times\nonumber\\
\left.\left(1-\frac{1}{2\log \frac{\tilde{U}}{\sqrt{\pi}}}-\frac{1}{4\log^2\frac{\tilde{U}}{\sqrt{\pi}}}+\ldots\right)\right].
\end{align}
A comparison between the stationary phase estimate and the actual numerical integral is graphed in Fig. \ref{fig:excitation_density}. 

\begin{figure}
 \centering
 \includegraphics[width=1.00\columnwidth]{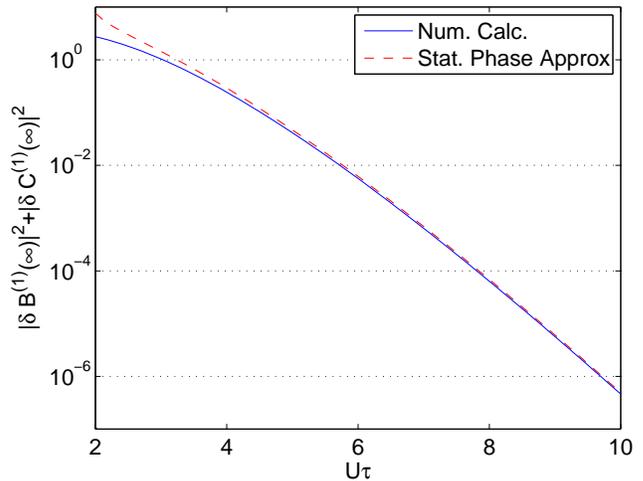}
 \caption{(Color online.) Leading order coefficient $\vert\delta B^{(1)}(\infty)\vert^2+\vert\delta C^{(1)}(\infty)\vert^2$ of the excitation density $\rho_{\textrm{exc}}$ in powers of $t_1^2/U^2$ plotted versus $U\tau$ for the specific case of a Gaussian pulse; see Eqs.~(\ref{eqnarray:excitation_density_form}) and (\ref{eqnarray:excitation_density}). The stationary phase estimate (dashed red) of the coefficient is a very good approximation to the full numerical calculation (solid blue) for $U\tau\gg1$. Both curves are formally accurate only for large $U\tau$.}
 \label{fig:excitation_density}
\end{figure}

To conclude this section, we note that there are many similarities between the $N$-site problem and the two-site problem. This is natural as we have expanded the Hamiltonian to only leading order in $t_1/U$, essentially linking a given site only with its nearest neighbors. Further neighbor terms in the Hamiltonian only appear at higher orders in $t_1/U$, and are correspondingly suppressed. We similarly expect that the limitations on the pulse fidelity coming from a competition between $\tilde{U}$ and $\kappa$ becomes important at $O(t_1^4/U^4)$.

\section{\label{sec:initstage}Initial Stage of Decay }
\begin{figure}
 \centering
 \includegraphics[width=0.80\columnwidth]{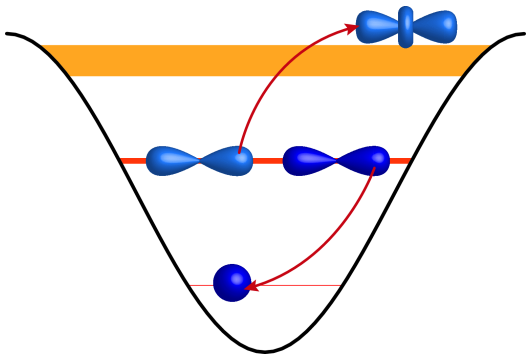}
 \caption{(Color online.) The principal on-site decay process in relatively shallow lattices ($\tilde{V}<26 E_{R}$) consists of two bosons in the $n=1$ band decaying into a localized $n=0$ particle and an a delocalized $n=2$ particle, leaving a localized hole in the $n=1$ band. For deep lattices, the decay process is prevented by the anharmonicity of the energy levels.}
 \label{fig:anharmonicity}
\end{figure}

\begin{figure}
 \centering
 \includegraphics[width=1.00\columnwidth]{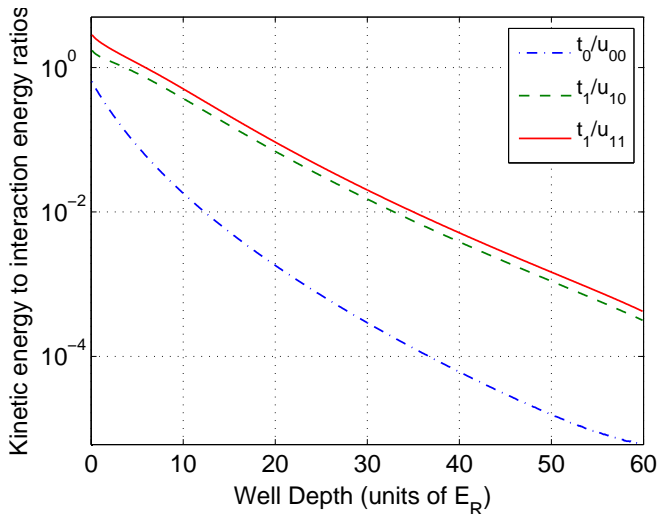}
 \caption{(Color online.) Ratios of the nearest neighbor hopping energy in bands $n=0,1$ to relevant on-site interaction energies plotted versus well depth.}
 \label{fig:t_vs_u}
\end{figure}

The \textit{p}-orbital insulator will decay since it is coupled to a continuum; nevertheless, its metastability is guaranteed by the anharmonicity of the optical lattice potential as shown in Fig. \ref{fig:anharmonicity}. For typical experimental parameters, the dominant decay mechanism roughly consists of an Auger process where two bosons in the $n=1$ band decay into a nearly free particle in the $n=2$ band and a localized $n=0$ particle at the site of interaction. We calculate this initial decay rate using Fermi's Golden Rule. Throughout this section, we are interested only in sufficiently deep well depths that $t_1/U\ll1$ and the width of the lowest band can be completely neglected. See Fig. \ref{fig:t_vs_u} for a plot of relevant ratios of the hopping energy in the ground and first excited bands $t_0,t_1$ to the on-site interaction energies $U_{00},U_{10}$ and $U_{11}$. For lattices deeper than $26 E_R$, the width of the $n=2$ band is too narrow to compensate for the anharmonicity of the potential and this simple Auger process is energetically forbidden. Instead, a more complicated three-body process mediated by a virtual intermediate state becomes important. 

\subsection{Multi-band Bose-Hubbard Hamiltonian}
Bosons loaded into an optical lattice are approximately described by the multi-band Bose Hubbard model (MBH) \cite{Isacsson05}. As we want to simulate the dynamics of bosons promoted to the $p$ orbital, it is useful to divide the MBH Hamiltonian into intraband terms $H_{\textrm{intra}}$ that preserve band occupation and interband terms $H_{\textrm{inter}}$ which drive transitions between bands:
\begin{subequations}
\label{eq:multi_band_BH_ham}
\begin{align}
H_{\textrm{intra}}=&\sum_{n}\epsilon_{n}\sum_{j}b_{nj}^{\dag}b_{nj}+\sum_{n}t_{nn}\sum_{j}b_{nj}^{\dag}b_{nj-1}+\textrm{h.c.}\nonumber\\
&+\frac{1}{2}\sum_{m,n}U_{mn}^{mn}\sum_{j}b_{mj}^{\dag}b_{nj}^{\dag}b_{nj}b_{mj},\\
H_{\textrm{inter}}=&\sum_{m}\sum_{n\neq m}t_{mn}\sum_{j}b_{mj}^{\dag}b_{nj-1}+\textrm{h.c.}\nonumber\\
&+\frac{1}{2}\sum_{m,n}\sum_{k,l\neq m,n}U_{mn}^{kl}\sum_{j}b_{kj}^{\dag}b_{lj}^{\dag}b_{mj}b_{nj},
\end{align}
\end{subequations}
with $t_{mn}$ being the hopping matrix element from band $m$ to band $n$ and $U_{mn}^{kl}$ being the interaction matrix element which drives pairs of bosons from bands $k,l$ to bands $m,n$. We have neglected further neighbor hoppings, which are small for the bands relevant to our calculation. To simplify notation, we use shorthand forms for the diagonal terms $t_n\equiv t_{nn}$ and $U_{mn}\equiv U_{mn}^{mn}$. The off-diagonal tunneling terms $t_{mn}$ only appear in perturbation theory with energy denominators $O(\epsilon_m-\epsilon_n)$; for the lattice parameters we are considering these are small and can be neglected. The system is in a sufficiently dilute limit that we can restrict our analysis to \textit{s}-wave scattering $a_{s}/\lambda\ll1$ \cite{Leggett2001}. The matrix elements $U_{mn}^{kl}$ are given by the expression:
\begin{subequations}
\begin{align}
U_{mn}^{kl}=g\int_{-\infty}^{\infty}W_k^*(z)W_l^*(z)W_m(z)W_n(z) dz,\\
g=16\pi^{2}\left(\frac{a_{s}}{\lambda}\right)E_R\left(\int_{-\infty}^{\infty}W_{0}^{\textrm{Vtrans}}(z)^{4}dz\right)^{2},
\end{align}
\end{subequations}
where $\lambda$ is the wavelength of the optical lattice lasers, $W_m$ are the Wannier functions for the lattice potential, and $W_{0}^{\textrm{Vtrans}}$ is the ground state Wannier function associated with the tightly confined transverse directions.  We will work using the scaled position variables $z=\frac{2\pi}{\lambda}x$ and scaled momentum variables $k=\frac{\lambda}{2 \pi}p$; in these variables, the first Brillouin zone covers the range $-1<k<1$. 

The nearest-neighbor hopping energy $t_{m}$ and interaction matrix element $U_{kl}^{mn}$ obey the following scaling relations:
\begin{subequations}
\begin{align}
\frac{t_{mn}}{E_R}=&\tilde{f}_{mn}\left(\frac{V}{E_R}\right),\\
\frac{U_{kl}^{mn}}{E_R}=&\tilde{g}_{mn}\left(\frac{V}{E_R},\frac{a_s}{\lambda}\right),
\end{align}
\end{subequations}
where $\tilde{f}_{mn}$ and $\tilde{g}_{mn}$ are system-independent functions. For the experimental parameters in \cite{Muller2007}, $a_s/\lambda=0.069\ldots$; we use this value for all subsequent calculations.

For the following discussion, we introduce second quantized operators $b_{nj}^{\dag}$, which create a localized particle in band $n$ at site $j$, and $b_{nk}^{\dag}$, which create a Bloch wave state in band $n$ with quasimomentum $k$. We exclusively use $j,j'$ for site indices, and $k,q,k',q'$ for momentum indices.
\subsection{Energetics}
To compute the short time decay rate of the $p$-orbital insulator, we calculate interband transitions between eigenstates of $H_{\textrm{intra}}$ using Fermi's Golden Rule; see Eq.~(\ref{eq:multi_band_BH_ham}). To leading order in $t_1/U$, the $p$-orbital $\left|\psi_0\right\rangle$ insulator state is given in second quantized notation by:
\begin{equation}
\left|\Psi_{0}\right\rangle \approx\left[1-\frac{t_{1}}{U_{11}}\sum_{\left\langle i'j' \right\rangle}\left(b_{1i'}^{\dag}b_{1j'}+b_{1j'}^{\dag}b_{1i'}\right)\right]\prod_{j}b_{1j}^{\dag}\left|0\right\rangle, 
\label{eq:init_state}\end{equation}
where $\left\langle i'j' \right\rangle$ indicates a sum over nearest neighbors. For experimentally relevant parameters, the $p$-orbital insulator can directly decay into only two energetically permissible final states, which we denote $|\psi_{Af}\rangle$ and $|\psi_{Bf}\rangle$. In terms of the second quantized operators $b_{nj}^{\dag}$ and $b_{nk}^{\dag}$, these states are approximately given by:
\begin{subequations}
\label{eq:psi_final}
\begin{align}
\left|\psi_{Af}(j,k,q)\right\rangle \approx b_{0j}^{\dag}b_{2q}^{\dag}b_{1k}b_{1j}\left|\psi_0\right\rangle,\\
\left|\psi_{Bf}(j,k,k',q)\right\rangle \approx b_{0j}^{\dag}b_{2q}^{\dag}b_{1k}b_{1k'}\left|\psi_0\right\rangle.
\end{align}
\end{subequations}
The difference between $|\psi_{Af}\rangle$ and $|\psi_{Bf}\rangle$ is the nature of the holes in the $n=1$ band: $|\psi_{Af}\rangle$ has an $n=1$ hole localized around the site of the interband transition, whereas $|\psi_{Bf}\rangle$ has two delocalized holes.  A detailed calculation gives that Decay Channel 1, where our initial state decays into $|\psi_{Af}\rangle$, is allowed energetically for $0<\tilde{V}<25.97$ and Decay Channel 2, where our initial state decays into $|\psi_{Bf}\rangle$, is allowed for $0<\tilde{V}<30.93$. The rates vanish above these threshold values of well depth.

In Appendix \ref{sec:r1r2_model}, we demonstrate how $\langle\psi_{Bf}\vert H_{\textrm{inter}}\vert\psi_0\rangle$ is suppressed by $O(t_1/U)$ compared to $\langle\psi_{Af}\vert H_{\textrm{inter}}\vert\psi_0\rangle$ by solving a representative model. This result is a consequence of the $n=1$ hole in $\vert\psi_{Af}\rangle$ being localized around the decay site $j=0$. The state $\vert\psi_{Bf}\rangle$ has a small $O(t_1/U)$ overlap with the decay site since it is orthogonal to  $\vert\psi_{Af}\rangle$.  This difference in the overlap between final states and the decay site leads to the difference in the size of the matrix elements. We neglect $R_2$, the decay rate corresponding to Decay Channel 2, since it is $O(t_1/U)^2$ smaller than $R_1$, the decay rate corresponding to Decay Channel 1. To simplify notation, we simply refer to $R_1$ as the initial decay rate $\Gamma$.

\subsection{Fermi's Golden Rule Calculation }
We now compute the principal decay rate $R_1$ using the full Hamiltonian in Eq.~(\ref{eq:multi_band_BH_ham}).  The final state is $\left|\psi_{Af}(j,k,q)\right\rangle$ defined in Eq.~(\ref{eq:psi_final}). The transition matrix element $U_{11}^{02}$ depends on both the quasimomentum of the $n=1$ band hole $k$ and the quasimomentum of the $n=2$ band particle $q$: 
\begin{equation}
U_{11}^{02}(k,q)=g\int_{-\infty}^{\infty}dz\ e^{i(k-q)z}u_{2q}(z)u_{1k}^{*}(z)W_{0}(z)W_{1}(z),
\end{equation}
where $u_{nk}\ $are the Bloch functions and $W_{n}(x)$ are the associated
Wannier functions \cite{Morsch2006}. The matrix element between the initial and final state is 
\begin{align}
\left\langle \psi_{f}(j,k,q)\left|H_{\textrm{inter}}\right|\psi_{0}\right\rangle&=\nonumber\\
&\frac{-4t_{1}U_{11}^{02}(k,q)e^{i\pi j(k-q)}\cos\left(k\pi\right)}{NU_{11}}.
\end{align}
We treat the interaction between the delocalized $n=2$ boson and the $p$-orbital insulator using a mean field approximation. This gives an additional mean field shift to the dispersion in the $n=2$ band:
\begin{subequations}
\begin{align}
\tilde{\epsilon}_{2q}=&\epsilon_{2q}+U_{12}(q),\\
U_{12}(k)=&g\int_{-\infty}^{\infty}dz\left|\ u_{2k}(z)\right|^{2}\sum_{j=-\infty}^{\infty}\left|W_{1}(z-j\pi)\right|^{2}.
\end{align}
\end{subequations}
The dependence of $U_{12}(k)$ on $k$ is weak and can be approximated by $U_{12}^{12}$. Substituting these terms into the Fermi Golden Rule formula for the initial decay rate gives \cite{Sakurai1993}
\begin{align}
\label{eq:full_rate}
\Gamma=\frac{32\pi}{\hbar}t_{1}^{2}\int_{0}^{1}dk\cos^{2}\left(\pi k\right)\int_{0}^{1}dq
\left|\frac{U_{11}^{02}(k,q)}{U_{11}}\right|^{2}\times\nonumber\\
\delta\left(\tilde{\epsilon}_{2q}+\epsilon_{0}-\epsilon_{1}-\epsilon_{1k}\right).
\end{align}

\begin{figure}
 \centering
 \includegraphics[width=1.00\columnwidth]{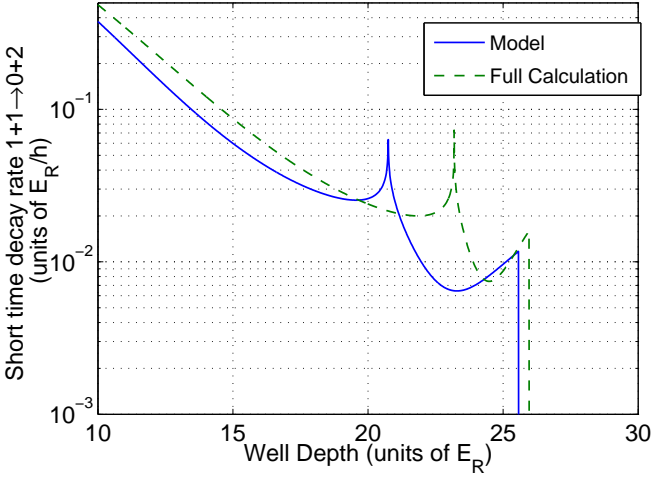}
 \caption{(Color online.) Comparison of the model short time rate with a full numerical calculation of the rate for the $p$-orbital insulator to decay to $\left|\psi_{Af}\right\rangle$; see Eqs.~(\ref{eq:psi_final}), (\ref{eq:full_rate}) and (\ref{eq:model_short_rate}). The model captures the presence of a logarithmic divergence in the rate due to a Van Hove singularity, as well as a jump discontinuity at threshold. }
 \label{fig:rate_a_comparison}
\end{figure}

\begin{figure}
 \centering
 \includegraphics[width=1.00\columnwidth]{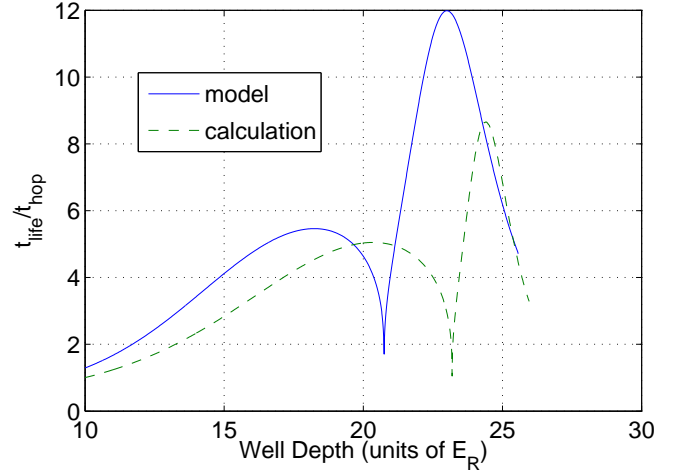}
 \caption{(Color online.) The short time lifetime $t_{\textrm{life}}$ in units of the hopping time $t_{\textrm{hop}}$ for both the full and model systems; see Eqs.~(\ref{eq:full_rate}) and (\ref{eq:model_short_rate}) and Fig. \ref{fig:rate_a_comparison}. }
 \label{fig:full_tlife_thop}
\end{figure}

To get insight into Eq.~(\ref{eq:full_rate}), we restrict band hoppings only to nearest neighbors, a valid limit for deep lattices. This approximation makes it possible to evaluate the decay rate $\Gamma$ analytically. As we show here, the rate predicts a jump discontinuity to zero above the threshold well depth. Keeping only nearest-neighbor hoppings, the dispersions in bands $n=1,2$ are given by:
\begin{subequations}
\label{eq:band_dispersion_simp}
\begin{align}
\tilde{\epsilon}_{2k}&\approx\tilde{\epsilon}_{2}-2\tilde{t}_{2}\cos(\pi k),\\
\epsilon_{1k}&\approx\epsilon_{1}+2t_{1}\cos(\pi k).
\end{align}
\end{subequations}
Note $\tilde{\epsilon}_{2k}$ includes the mean field shift. We treat the $n=1$ band hopping energy $t_1$ as much smaller than the effective $n=2$ band hopping energy $\tilde{t}_2$; this is consistent with experimental values. Using the simplified band dispersions from Eq.~(\ref{eq:band_dispersion_simp}), the expression for the short time decay rate $\Gamma$ is
\begin{align}
\label{eq:model_short_rate}
 \Gamma=\frac{16 t_{1}^{2}}{\hbar \tilde{t}_2}\int_{0}^{1}dk\frac{\left|U_{11}^{02}(k,q)\right|^{2}\cos^{2}\left(\pi k\right)}{\left(U_{11}\right)^2\sqrt{1-\left(A-B\cos\pi k\right)^2}}\times\nonumber\\
\theta\left(1-\left|A-B\cos\pi k\right|\right),
\end{align}
with the parameters $\lbrace A,B\rbrace$ only determined by well depth:
\begin{subequations}
\begin{align}
A(V)=&\frac{\tilde{\epsilon}_{2}+\epsilon_{0}-2\epsilon_{1}}{2\tilde{t}_{2}},\\
B(V)=&t_{1}/\tilde{t}_{2}.
\end{align}
\end{subequations}
The rate $\Gamma$ vanishes above a threshold value $V_t$, determined by the condition $|A(V_t)-B(V_t)|=1$. Below threshold, the rate can be approximated by Eq.~(\ref{eq:gamm_leading_order}) except near the well depth $V_{\textrm{sing}}$ that satisfies $|A(V_{\textrm{sing}})+B(V_{\textrm{sing}})|=1$. Here the rate diverges logarithmically due to the presence of a van Hove singularity. At threshold, the rate has a jump discontinuity to zero:
\begin{subequations}
\begin{align}
\Gamma(V_t^-)=&\frac{8}{\hbar}\frac{t_{1}^{3/2}\left|U_{11}^{02}\left(0,q\right)\right|^{2}}{\left(U_{11}\right)^2\tilde{t}_2^{1/2}},\\
\Gamma(V_t^+)=&0.
\end{align} 
\end{subequations}
The divergence in the rate and the jump discontinuity at threshold are also evident in the numerical evaluation of Eq.~(\ref{eq:full_rate}); see Figs. \ref{fig:rate_a_comparison} and \ref{fig:full_tlife_thop} for a comparison with the simplified model. A full discussion of the two-body decay rate near threshold is given in Appendix \ref{sec:threshold_behavior}. 

\subsection{Decay of \textit{p}-orbital Insulator above threshold}
Above threshold, the decay process to $\left|\psi_{Af}\right\rangle$ becomes energetically forbidden and one must go to higher order processes to determine the initial decay rate. Neglecting the effects of the magnetic trap, the dominant short time decay process is an effective three-body decay rate: 
\begin{align}
\label{eq:full_rate_3bod_decay_rate}
\Gamma=\frac{2\pi}{\hbar}\sum_{j\neq 0}\int_{-1}^{1}\frac{dk_1}{2} \int_{-1}^{1}\frac{dk_4}{2}\left|M\left(k_{1},k_{4},j\right)\right|^2\times\nonumber\\
\delta\left(
\tilde{\epsilon}_{4k_{4}}+2\epsilon_{0}-2\epsilon_{1}-\epsilon_{1k_{1}}\right),
\end{align}
where $\tilde{\epsilon}_{4k_{4}}$ includes the mean field shift. This rate is energetically allowed for $V\lessapprox 36E_R$. Although the second-order perturbation theory expression for the three-body interaction matrix element includes a sum on all intermediate bands, the contribution from the $n=2$ band is most important:
\begin{align}
M(k_{1},k_{4},j)\approx&\frac{-8t_1 \cos(\pi k_1)}{U_{11}}\nonumber\times\\
&\int_{-1}^{1}\frac{dk}{2}\frac{U_{12}^{04}(k_{4},k)U_{11}^{02}(k)e^{i\pi kj}}{\epsilon_{1}+\epsilon_{1k_{1}}-\epsilon_{2k}-\epsilon_{0}}.
\end{align}
A recent paper addressed an effective three-body decay mechanism utilizing an intermediate virtual state in a related system \cite{Mazets2008}. 
\subsection{Combining two- and three-body decay rates}
\begin{figure}
 \centering
 \includegraphics[width=1.00\columnwidth]{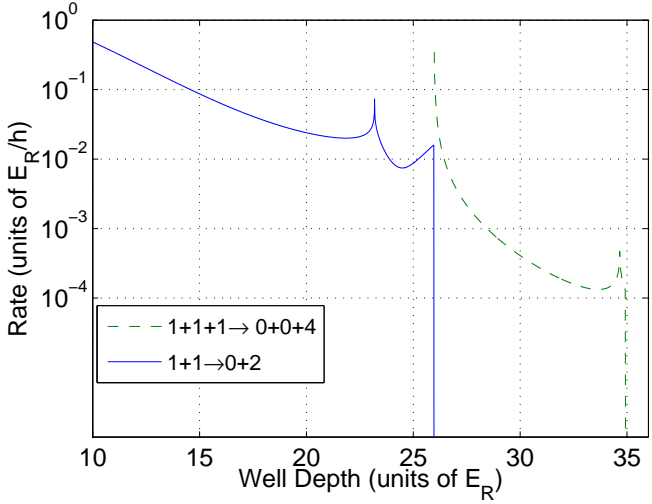}
 \caption{(Color online.) Comparison of the short time three-body decay rate $1+1+1\rightarrow0+0+4$ and the short time two-body decay rate $1+1\rightarrow0+2$; see Eqs.~(\ref{eq:full_rate}) and (\ref{eq:full_rate_3bod_decay_rate}). Note the two-body rate exhibits a discontinuous jump to zero at threshold, whereas the three-body rate diverges.}
 \label{fig:3bod_2bod_comp}
\end{figure}

\begin{figure}
 \centering
 \includegraphics[width=1.00\columnwidth]{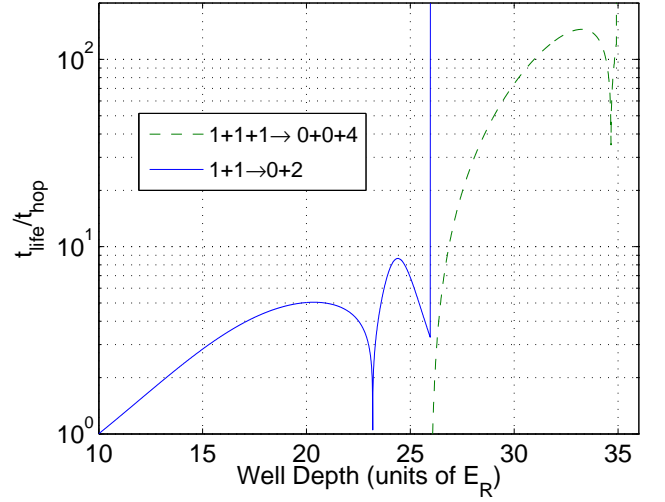}
 \caption{(Color online.) Comparison of the short time three-body and two-body decay lifetimes in units of the hopping time $t_{hop}=2t_1/h$; see Eqs.~(\ref{eq:full_rate}) and (\ref{eq:full_rate_3bod_decay_rate}) and Fig. \ref{fig:3bod_2bod_comp}.}
 \label{fig:3bod_2bod_hop}
\end{figure}

We expand the behavior of both the two- and three-body short time decay rates near the threshold value $V_t$ in Appendix \ref{sec:threshold_behavior}. For $V<V_t$, we can approximately describe the early decay dynamics by just the two-body decay rate; see Eq.~(\ref{eq:full_rate}). As $V\rightarrow V_t^-$, we find that the two-body rate rises linearly to a constant before jumping to zero: 
\begin{subequations}
\begin{align}
\label{eq:two_bod_threshold}
\Gamma(V<V_t)\approx&\frac{8\left|U_{11}^{02}(1)\right|^2t_1^{3/2}}{\hbar U_{11}^2\tilde{t}_2^{1/2}}\times\nonumber\\
&\left[1-\left.\frac{7}{16}\frac{d\overline{g}(\tilde{V})}{d\tilde{V}}\right|_{V_t}\left(\frac{V-V_t}{t_1}\right)\right],\\
\overline{g}(\tilde{V})=&\frac{1}{E_R}\left(\tilde{\epsilon}_{2}+2\tilde{t}_{2}+\epsilon_{0}-2\epsilon_{1}+2t_{1}\right),
\end{align}
\end{subequations}
where $\tilde{V}$ is the well depth in units of recoil energy. For well depths above the two-body threshold ($V>V_t$) we can describe the dynamics using the effective three-body decay rate; see Eq.~(\ref{eq:full_rate_3bod_decay_rate}). As $V\rightarrow V_t^+$, the three-body rate becomes singular as one approaches threshold:
\begin{equation}
\label{eq:three_bod_threshold}
 \Gamma\approx\frac{8\left(t_1\right)^{3/2}\left\vert U_{11}^{02}(1)\right\vert^2\left\vert U_{12}^{04}(1,k_4^0)\right\vert^2}{\hbar\sqrt{t_2}U_{11}^2\left\vert\frac{d\epsilon_{4k}}{dk}\right\vert_{k_4^0}\left[-\left.\frac{d}{d\tilde{V}}\overline{g}(\tilde{V})\right\vert_{V_t}(V-V_t)\right]}.
\end{equation}
This singularity can be understood in the following way: the virtual intermediate state is progressively longer lived for well depths closer and closer to threshold. At threshold the state becomes truly long-lived and there is a divergence in the decay rate. Comparisons of the short time two-body decay rate $1+1\rightarrow0+2$ and the three-body decay rate $1+1+1\rightarrow0+0+4$ are plotted in Figs. \ref{fig:3bod_2bod_comp} and \ref{fig:3bod_2bod_hop}.

\begin{figure}
 \centering
 \includegraphics[width=1.00\columnwidth]{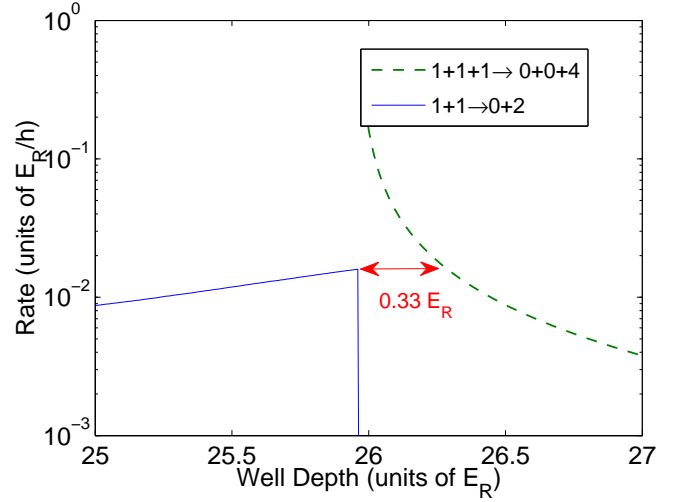}
 \caption{(Color online.) Comparison of the short time three-body and two-body decay rates near the threshold value $V_t$; see Eqs.~(\ref{eq:two_bod_threshold}
) and (\ref{eq:three_bod_threshold}). Just below threshold, the two-body decay rate reaches a value of about $0.016 E_R/h$, after which it jumps to zero. The three-body rate reaches this same value at a well depth only $0.33E_R$ above threshold. We expect that the Golden Rule approximation breaks down only in a small region around threshold, with a width of about $E_R$ on either side.}
 \label{fig:3bod_2bod_threshold}
\end{figure}

This approach only fails for a small region near threshold, where the analysis is complicated by the presence of an allowed real intermediate state for $V<V_t$. To benchmark the width of this inapplicable region, we calculated the well depth $V_{\textrm{3bod}}$ where the three-body decay rate equals the two-body rate just below threshold (see Fig. \ref{fig:3bod_2bod_threshold}). We found $V_{\textrm{3bod}}-V_t=0.33 E_R$, a small value compared to the experimentally accessible range of well depths. 

\begin{figure}
 \centering
\includegraphics[width=1.00\columnwidth]{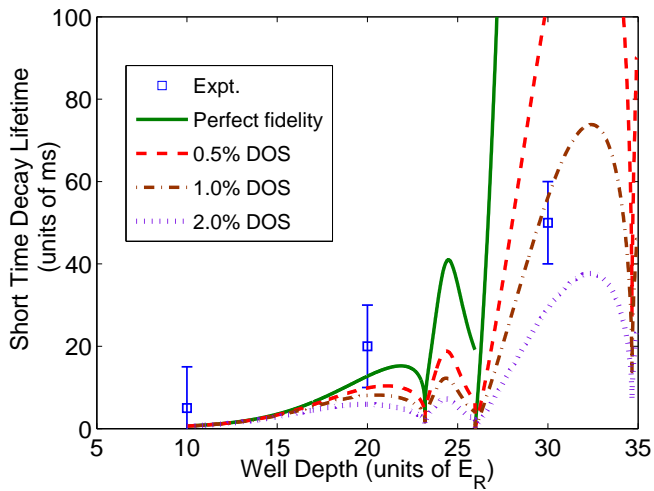}
\caption{(Color online.) Comparison of our theoretical short times rate with lifetimes extracted from experimental measurements published by M\"uller et al. \cite{Muller2006}. We have plotted the lifetime for the $p$-orbital insulator, as well as ``imperfect'' excited states with an additional $0.5\%,1.0\%$ and $2.0\%$ fraction of double occupied sites (DOS). The measured values appear consistent with roughly $1\%$ DOS. Sources of these excitation infidelities could come from imperfect $\pi$-pulses.}
\label{fig:short_times_expt}
\end{figure}

Linking the short time rates to the measurements made by M\"uller et al. is made difficult because our approach neglects the spreading of atoms in the trap and the effect of an imperfect $\pi$-pulse \cite{Muller2006}. We can, however, define a rough time scale for the experiments where the optical lattice system crossed over from early stage dynamics to late stage dynamics. Fitting the experimental data with a double exponential, we can define the inverse of the faster decay rate to be the ``short time decay lifetime''. For the well depths considered, this lifetime was typically tens of milliseconds. In Fig. \ref{fig:short_times_expt} we compare this observed lifetime with the Fermi Golden Rule rate for the $p$-orbital insulator. Our predicted short time rates are of the same order as those observed in experiment for well depths $V=10E_R$ and $20E_R$, but they differ substantially for $V=30E_R$. 

\begin{figure}
 \centering
\includegraphics[width=1.00\columnwidth]{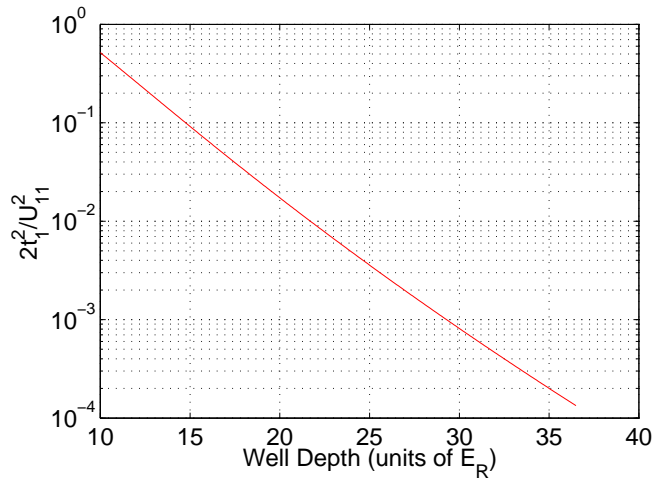}
\caption{(Color online.)The $p$-orbital insulator has roughly a $\frac{2t_1^2}{U_{11}^2}$ fraction of double occupied sites. For $V\lesssim 20E_R$, this fraction is larger than the estimated fraction of sites doubly occupied due to pulse infidelities in experiment \cite{Muller2007}, and we expect that the short time decay rate of the experimental system will be similar to that of the $p$-orbital insulator. For deeper well depths $V>20E_R$, we expect that the short time decay rate is dominated by decays of sites doubly occupied due to pulse infidelities. }
\label{fig:ratio_2t1_sq_over_u11_sq}
\end{figure}

One possible explanation for this difference with experimental data is that the system has a small fraction of double occupied sites due to an imperfect preparation. Neglecting interference terms, the Golden Rule rate is roughly the fraction of double occupied sites multiplied by the on-site decay rate of a single site. As shown in Fig. \ref{fig:short_times_expt}, the observed data seems to be consistent with about $1\%$ of sites doubly occupied due to preparation imperfections. Since the $p$-orbital insulator alone has roughly $2t_1^2/U_{11}^2$ double occupied sites due to virtual hops, the short time decay rate of the experimental system will substantially differ from theory only when the fraction of ``accidentally'' double-occupied sites becomes comparable to $2t_1^2/U_{11}^2$. As shown in Fig. \ref{fig:ratio_2t1_sq_over_u11_sq}, the ratio $2t_1^2/U_{11}^2>0.01$ for $V\lesssim 22E_R$, but rapidly falls below $0.01$ for deeper well depths. Reducing the number of double occupied sites through improved $\pi$-pulses could substantially improve excited band lifetimes, particularly for well depths $V>20E_R$. 

\section{\label{sec:thermeq}Thermal equilibrium }

\begin{figure}
 \centering
 \includegraphics[width=1.00\columnwidth]{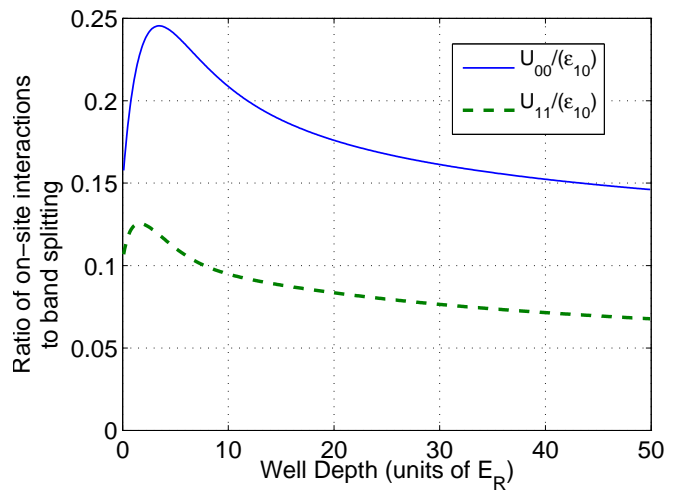}
 \caption{(Color online.) Plot of on-site interaction energies versus band splitting. As the equilibrium temperature of the system is $O(\epsilon_{10})$, we can approximately neglect interactions and treat the final state as a free Bose gas. }
 \label{fig:onsite_vs_splitting}
\end{figure}

After the initial stage of decay, the $p$-orbital insulator thermalizes and spreads in the trap. This stage of decay is difficult to model. At very long times the distribution is described by Bose-Einstein statistics, with the chemical potential $\mu$ and the temperature $T^e$ determined by total energy and particle number. The interactions among the particles (of order $U$) are much smaller than the temperature (of order the band splitting $\epsilon_{10}$); see Fig. \ref{fig:onsite_vs_splitting}.

\begin{figure}
 \centering
 \includegraphics[width=1.00\columnwidth]{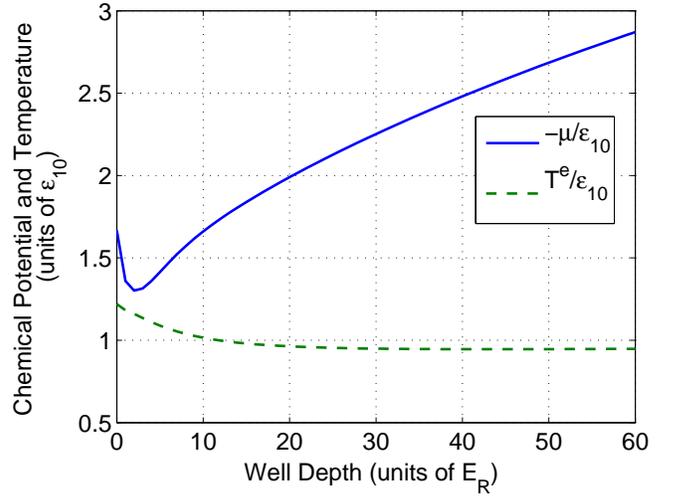}
 \caption{(Color online.) The solution for the chemical potential $\mu$ and equilibrium temperature $T^e$ including mean field shifts from Eq.~(\ref{eq:energy_bose}). Here we have plotted $-\mu/\epsilon_{10}$ in order to easily compare the two graphs. Note the y-axis starts from 0.5.}
 \label{fig:chem_pot_full_with_mf}
\end{figure}

\begin{figure}
 \centering
 \includegraphics[width=1.00\columnwidth]{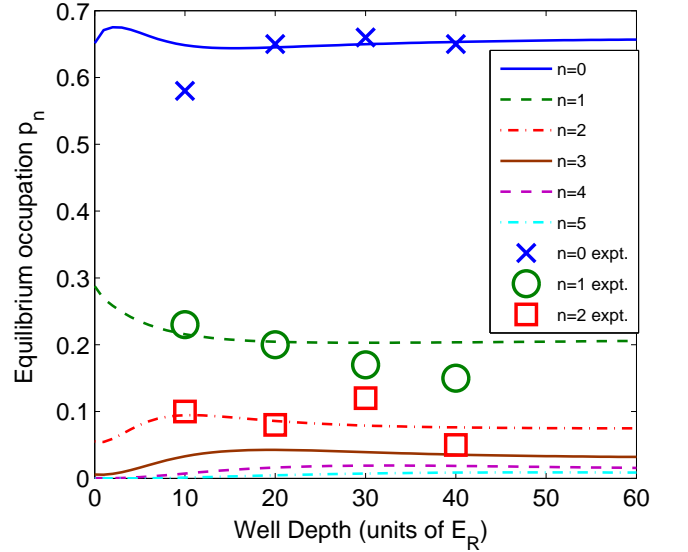}
 \caption{(Color online.) The full probability distribution including mean field shifts from Eq.~(\ref{eq:energy_bose}). The ``X'' markers represent the experimental measurement of the equilibrium values of bands $n=0,1,2$ \cite{Muller2006}. The agreement with experiment is reasonably good as expected from an equilibrium calculation with proper Bose statistics; possible differences between our experiment and theory could arise from neglecting excitations in the $y-$ and $z-$ directions. }
 \label{fig:full_probability_stats_with_mf}
\end{figure}

The magnetic trap reduces the equilibrium temperature $T^e$ in a manner analogous to the virial theorem, just as M\"uller et al. initially proposed \cite{Muller2006}. To simplify matters, we consider that the trap is initially loaded with bosons just up to the point where double occupancies are favorable. Here the condition on the number of particles is given by $N=2\sqrt{U_{00}/A}$, where $U_{00}$ is the on-site energy in the ground band and the trap potential is given by $A(2x/\lambda)^2$. 
As the length scale of the trap is so much longer than that of the optical lattice parameter we assume we can treat the dispersion semiclassically: 
\begin{equation}
\epsilon_{nk}(x)\approx\epsilon_{nk}+4Ax^2/\lambda^2.
\end{equation}
We treat interactions in the mean field. As the temperature $T^e$ is much larger than the width of relevant bands, we ignore any momentum dependence of the interaction $U_{nn'}$ between particles in bands $n$ and $n'$.  Introducing the band relative occupancy $p_n=\frac{1}{N}\sum_k f_{nk}$, the equilibrium mean field shift $F_{n0}^e$ in band $n$ is
\begin{subequations}
\begin{align}
&F_{n0}^e=\rho_0\sum_{n'}\rho_{m}'U_{nm}p_{m},\\
&\rho_n'=\sqrt{\frac{4U_{00}}{\pi T^e}}\frac{\sum_k\left\{\exp\left[\left(\epsilon_{nk}+F_{n}-\mu\right)/T^e\right]-1\right\}^{-1}}{\sum_k\textrm{Li}_{1/2}\left\{\exp\left[-\left(\epsilon_{nk}+F_{n}-\mu\right)/T^e\right]\right\}},
\end{align}
\end{subequations}
with Li$_k$ the polylogarithm. The $p_n$ are determined self-consistently. Using these approximations, the equations for number, energy and relative occupancy $p_n$ become
\begin{subequations}
\label{eq:energy_bose}
\begin{align}
&\sum_np_n=1,\\
&p_{n}=\frac{1}{N}\sum_{k}\int_{-\infty}^{\infty}\frac{dx}{\lambda/2}\frac{1}{\exp\left[\frac{\epsilon_{nk}(x)+F_{n0}^e-\mu}{T}\right]-1},\\
&\frac{1}{N}\sum_{nk}\int_{-\infty}^{\infty}\frac{dx}{\lambda/2}\frac{\epsilon_{nk}(x)}{\exp\left[\frac{\epsilon_{nk}(x)+F_{n0}^e-\mu}{T}\right]-1}=\epsilon_{1}+\frac{U_{00}}{3}.
\end{align}
\end{subequations}
The values for the equilibrium temperature $T^e$ and chemical potential $\mu$ obtained from these implicit equations are plotted in Fig. \ref{fig:chem_pot_full_with_mf}. The relative occupancies $p_{0-5}$ are compared to experimental results from \cite{Muller2006} in Fig. \ref{fig:full_probability_stats_with_mf}; some differences between our theory and experimentally measured values are expected to come from neglecting excitations in the $y-$ and $z-$ directions.

\section{\label{sec:longtime}Long Time Decay}
We want to model the long time relaxation to the equilibrium computed in the previous section through a Boltzmann equation formalism. The full Boltzmann equation includes both spatial terms which lead to a spreading in the trap and momentum terms which equilibrate the different bands. To make our analysis mathematically tractable we ignore the spatial terms, even though we suspect that they are important in understanding experimental results. We treat the density within the trap as fixed by its equilibrium value and allow thermalization only through momentum and interband relaxation. Rather than a complete theory of the later stages of evolution, this section should be considered a first attempt to model the dynamics of interband scattering. 

Initially the trap is loaded just to the point where double occupancies become energetically favorable. As discussed in section \ref{sec:thermeq}, this maximally loaded condition implies $N=2\sqrt{U_{00}/A}$. We derive the Boltzmann Equation first assuming classical statistics in Section \ref{subsec:classical} and then for Bose statistics in Appendix \ref{sec:bose_derive}. We solve these two equations numerically in Section \ref{subsec:eigvec}.

\subsection{\label{subsec:classical}Derivation of Boltzmann Equation with Classical Statistics}
For classical statistics to be valid we require the occupation of each band to be relatively small. This condition fails for the lowest bands that have relatively high occupation. We relax this condition in Appendix \ref{sec:bose_derive} when we derive the Boltzmann equation with Bose statistics.

We work in the semiclassical approximation where we can specify the occupation $f_{nkj}$ of band $n$, quasimomentum $k$, and site $j$. Although this seems at odds with the uncertainty principle, it is a valid approximation when the wavelength of the optical lattice $\lambda/2$ is much shorter than the band-dependent harmonic oscillator length $\xi_n\sim \left(\hbar\lambda\right)^{1/2}\left(Am_n^{\textrm{eff}}\right)^{-1/4}$, with $m_n^{\textrm{eff}}$ the characteristic mass scale in band $n$ defined by the bandwidth: $\hbar^2/m_n^{\textrm{eff}}\lambda^2\sim t_n$. For the case of maximal loading, the ratio $\xi_n/\lambda$ is roughly
\begin{equation}
 \frac{\xi_n}{\lambda}\sim\left(\frac{t_n}{A}\right)^{1/4}.
\end{equation}
For experimental parameters, $\xi_n/\lambda\gg1$ for all bands but $n=0$. For deeper lattice well depths, $\xi_0/\lambda \gtrsim 1$ suggesting the semiclassical approximation may not be valid experimentally for the lowest band. In the following, we make the simplification that the trap is sufficiently weak to apply the semiclassical approximation in every band. The semiclassical energy $\epsilon_{nkj}$ of a particle in band $n$, quasimomentum $k$ and site $j$ including the mean field shift $F_{nj}\equiv\frac{1}{N}\sum_{n'k'}U_{nn'}f_{n'k'j}$ is
\begin{equation}
 \epsilon_{nkj}\approx\epsilon_{nk}+Aj^2+F_{nj}.
\end{equation}
Since the $F_{nj}$ are small compared to the equilibrium temperature $T^e$, we can ignore the $j-$ dependence and use $F_{nj}\approx F_{n0}$. The classical distribution of $f_{nkj}$ at thermal equilibrium is given by: 
\begin{equation}
f_{nkj}^e=N\frac{\exp\left(-\epsilon_{nkj}/T^e\right)}{\frac{1}{N}\sum_{nkj}\exp\left(-\epsilon_{nkj}/T^e\right)},
\end{equation}
where the equilibrium temperature $T^e$ is determined by conservation of energy. For the case of a maximally loaded trap this is given by:
\begin{equation}
\label{eq:en_cons_full}
 \frac{1}{N}\sum_{nkj}\epsilon_{nkj}f_{nkj}^e=N\left(\epsilon_1+\frac{U_{00}}{3}\right),
\end{equation}
which simplifies to
\begin{align}
\label{eq:en_cons_trap}
\epsilon_1+\frac{U_{00}}{3}&=\nonumber\\
&\frac{T^e}{2}+\frac{\sum_{nk}\left(\epsilon_{nk}+F_{n0}^e\right)\exp\left[-\left(\epsilon_{nk}+F_{n0}^e\right)/T^e\right]}{\sum_{nk}\exp\left[-\left(\epsilon_{nk}+F_{n0}^e\right)/T^e\right]},
\end{align}
where $F_{n0}^e$ is the equilibrium mean field shift. The $\frac{T^e}{2}$ in Eq.~(\ref{eq:en_cons_trap}) is the virial term which represents the energy contribution from the trap potential. This term was first suggested to explain experimental data \cite{Muller2007}. Neglecting all spatial derivatives, the Boltzmann equation becomes
\begin{align}
\label{eq:boltz_full_v1}
\frac{df_{n_1k_1j}}{dt}=-\sum_{n_2}\sum_{n_3\geq n_4}\frac{1}{N^3}\sum_{k_{2},k_3,k_4}
\Gamma_{n_1n_2j}^{n_3n_4}(k_1,k_2,k_3,k_4)\times\nonumber\\
\left(f_{n_1k_1j}f_{n_2k_2j}-f_{n_3k_3j}f_{n_4k_4j}\right),
\end{align}
with the function $\Gamma_{n_1n_2j}^{n_3n_4}$ given by:
\begin{align}
\label{eq:gamm_full}
\Gamma_{nn_2j}^{n_3n_4}(k_{1},k_2,k_3,k_4)=\frac{2\pi}{\hbar}\left|U_{n_1 n_{2}}^{n_{3}n_{4}}(k_{1},k_2,k_3,k_4)\right|^{2}\times\nonumber\\
\delta\left(k_1+k_2-k_3-k_4+G\right)\times\nonumber\\
\delta\left(\epsilon_{n_1k_1j}+\epsilon_{n_{2}k_{2}j}-\epsilon_{n_{3}k_{3}j}-\epsilon_{n_{4}k_{4}j}\right).
\end{align}
We now make the simplifying ansatz that within a given band the occupation $f_{nk0}$ is Boltzmann distributed, but the relative occupation of bands $p_n$ can fluctuate:
\begin{subequations}
\label{eq:ansatz_classical}
\begin{align}
&f_{nk0}\approx\rho_0p_n\frac{\exp\left[-\left(\epsilon_{nk}+F_{n0}\right)/T\right]}{A_n},\\
&\rho_0=\sqrt{\frac{4U_{00}}{\pi T^e}},\\
&A_n\equiv\frac{1}{N}\sum_k\exp\left[-\left(\epsilon_{nk}+F_{n0}\right)/T\right],
\end{align}
\end{subequations}
where $\rho_0$ is the density near the center of the trap and $T$ is an effective temperature determined by conservation. At equilibrium, the band occupancy probabilities $p_n$ and mean field shifts $F_{n0}$ are given by:
\begin{subequations}
\begin{align}
&p_n^e=A_n(T^e)/{\cal Z},\\
&F_{n0}^e=\rho_0\sum_{n'}U_{nn'}p_{n'}^e,
\end{align}
\end{subequations}
where the partition function ${\cal Z}=\sum_nA_n(T^e)$.
We allow the temperature to fluctuate in time in order to enforce conservation of energy. Using the ansatz in Eq.~(\ref{eq:ansatz_classical}) and summing Eq.~(\ref{eq:boltz_full_v1}) over $k_1$ gives an equation for the $p_n$'s: 
\begin{equation}
\label{eq:boltz_full}
 \frac{dp_{n_1}}{dt}=-\rho_0\sum_{n_2,n_3\geq n_4} \gamma_{n_1n_2}^{n_3n_4}(T)\left(\frac{p_{n_1}p_{n_2}}{A_{n_1}A_{n_2}}-\frac{p_{n_3}p_{n_4}}{A_{n_3}A_{n_4}}\right),
\end{equation}
with $\gamma_{n_1n_2}^{n_3n_4}(T)$ given by the expression:
\begin{align}
\label{eq:gamma_nn2_n3n4}
 \gamma_{n_1n_2}^{n_3n_4}(T)=\frac{1}{N^4}&\sum_{k_1k_2k_3k_4}\Gamma_{n_1n_20}^{n_3n_4}\left(k_{1},k_{2},k_{3},k_{4}\right)\times\nonumber\\
&\exp\left(-\frac{\epsilon_{n_1k_1}+F_{n_10}^e+\epsilon_{n_2k_2}+F_{n_20}^e}{T}\right).
\end{align}
Since the temperature $T$ is much larger than the widths of the bands we are considering, the exponential factor in Eq.~(\ref{eq:gamma_nn2_n3n4}) depends very weakly on momentum. For a given well depth, the coefficients $\gamma_{nn_2}^{n_3n_4}(T)$ are nonzero for the values of $n,n_2,n_3,n_4$ where the two-body scattering process $n+n_2\rightarrow n_3+n_4$ conserves energy; see Eq.~(\ref{eq:gamm_full}). When the well depth becomes sufficiently large that the scattering process $n+n_2\rightarrow n_3+n_4$ no longer conserves energy, $\gamma_{nn_2}^{n_3n_4}$ will jump from a nonzero value to zero. These jumps have a substantial effect on lifetimes in the excited band, as can be seen in Fig. \ref{fig:eigen_life_all}.

Expanding the Boltzmann equation [Eq.~(\ref{eq:boltz_full})] to first order in deviations in both $p_n$ and $T$ from equilibrium gives:
\begin{widetext}
\begin{align}
\label{eqnarray:lin_boltz_eq}
 \frac{d\delta p_{n_1}}{dt}=-\frac{\rho_0}{\cal Z}\sum_{n_2,n_3\geq n_4}\gamma_{n_1n_2}^{n_3n_4}\left\{\frac{\delta p_{n_1}}{A_{n_1}}+\frac{\delta p_{n_2}}{A_{n_2}}-\frac{\delta p_{n_3}}{A_{n_3}}-\frac{\delta p_{n_4}}{A_{n_4}}+\left[\frac{\rho_0}{{\cal Z}}\sum_m\delta p_m\left(\frac{U_{mn}+U_{mn_2}-U_{mn_3}-U_{mn_4}}{T^e}\right)\right]\nonumber \right.\\
\left.+\delta T\left[\frac{\left\langle \epsilon_{n_3}\right\rangle+F_{n_30}^e+\left\langle \epsilon_{n_4}\right\rangle+F_{n_40}^e-\left\langle\epsilon_{n_1}\right\rangle-F_{n_1}^e-\left\langle\epsilon_{n_2}\right\rangle-F_{n_20}^e}{\left(T^e\right)^2{\cal Z}}\right]\right\},
\end{align}
\end{widetext}
where $\langle\rangle$ indicates the average over a single band: 
\begin{equation}
 \langle g_n\rangle=\frac{\sum_{nk}g_{nk}\exp\left(-\epsilon_{nk}/T\right)}{\sum_{nk}\exp\left(-\epsilon_{nk}/T\right)}.
\end{equation}
Since Eq.~(\ref{eq:boltz_full}) does not mix $f_{nkj}$ on different sites we must have energy conservation on-site. This requirement is expressed by the relation
\begin{equation}
\label{eq:en_cons_site}
 \delta\sum_{nk}\left(\epsilon_{nk}+F_{n0}^e\right)f_{nk0}=0.
\end{equation}
Eq.~(\ref{eq:en_cons_site}) leads to the following equation for $\delta T$ in terms of the $\delta p_n$:
\begin{equation}
\label{eq:delta_T_eq}
\delta T=\frac{-\sum_{n}\delta p_n\left(\langle\epsilon_n\rangle+2F_{n0}^e\right)}{1/{\cal Z}\sum_nA_n(T^e)\left[\left(\langle\epsilon_n^2\rangle-\langle\epsilon_n\rangle^2\right)/\left(T^e\right)^2\right]}.
\end{equation}
We find the eigenmodes of Eq.~(\ref{eqnarray:lin_boltz_eq}) in Section \ref{subsec:eigvec}.

\subsection{\label{subsec:eigvec} Solution of Linearized Boltzmann Equations}
We computed the $\lbrace\gamma_{nn_2}^{n_3n_4}\rbrace$ with both classical and Bose statistics for nine bands over a range of well depths $0-40 E_R$; see Eqs.~(\ref{eq:gamma_nn2_n3n4}) and (\ref{eq:gamma_nn2_n3n4_bose}). Using these parameters, we determined the eigenvalues of the linearized Boltzmann equation for the cases of both classical [Eq.~(\ref{eqnarray:lin_boltz_eq})] and Bose statistics [Eq.~(\ref{eqnarray:lin_boltz_eq_bose})]; these eigenvalues represent the decay rates for each mode of the system. The linearized Boltzmann equation has two modes with zero eigenvalues corresponding to number and energy conservation. 

\begin{figure}
 \centering
 \includegraphics[width=1.00\columnwidth]{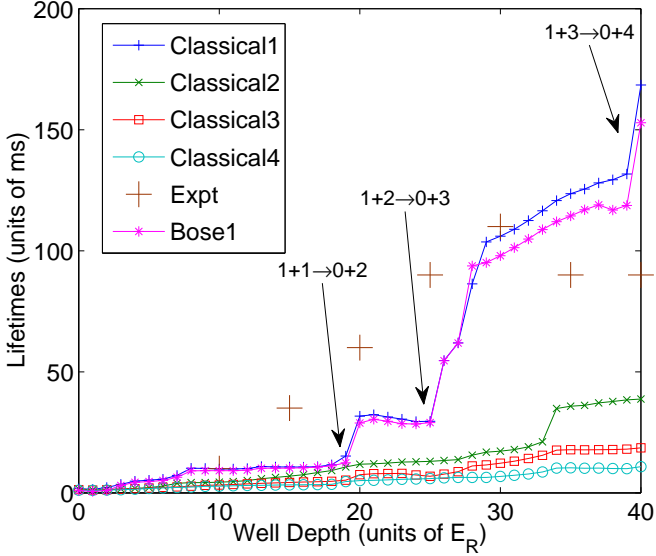}
 \caption{(Color online.) Comparison of the slowest-decaying modes of the linearized Boltzmann equation with both classical statistics [Classical 1-4] and Bose statistics [Bose1] with the slowest-decaying mode extracted from published measurements of the occupancy of band $n=1$ by M\"uller et al. [Expt.] \cite{Muller2007}; see Eqs.~(\ref{eqnarray:lin_boltz_eq}) and (\ref{eqnarray:lin_boltz_eq_bose}) . The inclusion of Bose factors gives only a small effect, so we have plotted only the slowest-decaying mode for that case. As the optical lattice potential becomes deeper, some decay channels become energetically forbidden and their decay rates jump to zero. As noted in the figure, crossing some of these thresholds is accompanied by a substantial increase in the lifetime of the excited band.}
 \label{fig:eigen_life_all}
\end{figure}

\begin{figure}
 \centering
 \includegraphics[width=1.00\columnwidth]{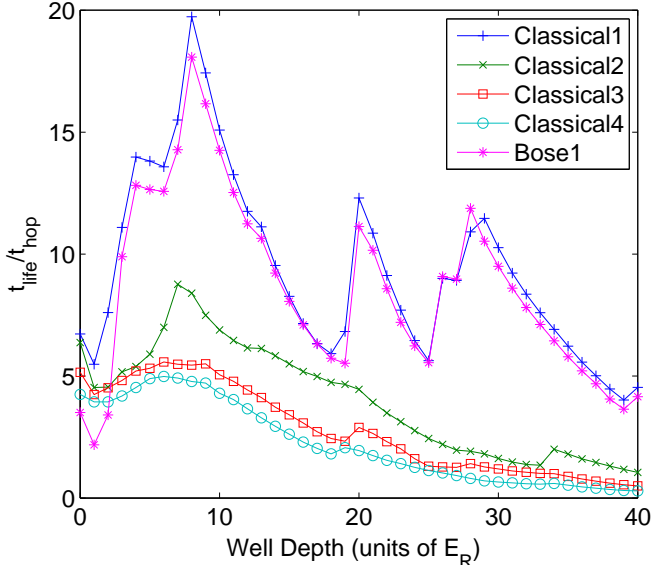}
 \caption{(Color online.) Lifetimes of the slowest-decaying modes of the linearized Boltzmann equation with both classical statistics [Classical1-4] and Bose statistics [Bose1] in units of the hopping time; see Eqs.~(\ref{eqnarray:lin_boltz_eq}) and (\ref{eqnarray:lin_boltz_eq_bose}). This dimensionless ratio is one of the most important figures of merit in characterizing the feasibility of studying quasi-equilibrium models in the first excited band.}
 \label{fig:eigen_hops_all}
\end{figure}

The lifetimes (1/rate) of the four slowest-decaying nonzero modes of the Boltzmann equation with classical statistics are plotted in Fig. \ref{fig:eigen_life_all}. The comparison between our numerical calculations and experimental measurements is relatively weak; this discrepancy could be due to neglecting the spatial dependence of the Boltzmann equation, which is known to be important in experiment but was neglected in our analysis. Inclusion of Bose factors gives only a small effect due to the high effective temperature $T$ and relatively low band occupation probabilities $p_n$. The lifetime of the slowest-decaying nonzero mode of the Boltzmann equation with Bose statistics is shown for comparison in Fig. \ref{fig:eigen_life_all}. We plot these lifetimes in units of the hopping time in Fig. \ref{fig:eigen_hops_all}. 

In Table \ref{tab:eigvector_bose} we decompose the slowest-decaying mode $v_1$ of the linearized Boltzmann equation with Bose statistics onto fluctuations in the different bands: $v_1=\sum_nv_{1n}\delta p_n$. The decomposition of the slowest-decaying mode of the linearized Boltzmann equation with classical statistics is nearly identical. For lattice well depths near $10E_R$ the slowest-decaying mode has little overlap with $\delta p_1$, suggesting that fluctuations from equilibrium in the first excited band relax faster than fluctuations in other bands at this value of well depth. 

\begin{table}
\centering
\begin{tabular}{|c|c|c|c|c|c|c|c|c|}
\hline 
$\tilde{V}(E_R)$&&$v_{10}$&$v_{11}$&$v_{12}$&$v_{13}$&
$v_{14}$&$v_{15}$&$v_{16}$\tabularnewline\hline\hline 
5.0&&0.624&-0.772&-0.034&0.110&0.031&0.005&0.000\tabularnewline\hline 
10.0&&0.601&-0.099&-0.775&-0.005&0.156&0.063&0.013\tabularnewline\hline 
15.0&&0.452&0.289&-0.727&-0.384&0.118&0.142&0.056\tabularnewline\hline 
20.0&&-0.478&0.854&-0.145&-0.129&-0.065&-0.013&-0.001\tabularnewline\hline 
25.0&&-0.488&0.850&-0.166&-0.098&-0.043&-0.021&-0.008\tabularnewline\hline 
30.0&&-0.530&0.832&-0.014&-0.142&-0.078&-0.027&-0.011\tabularnewline\hline 
35.0&&-0.532&0.833&-0.029&-0.130&-0.071&-0.027&-0.013\tabularnewline\hline 
40.0&&-0.545&0.826&-0.011&-0.109&-0.081&-0.029&-0.016\tabularnewline\hline 
\end{tabular}
\caption{Eigenvector decomposition of the slowest-decaying mode of the linearized Boltzmann equation with Bose statistics [Eq.~(\ref{eqnarray:lin_boltz_eq_bose})] on to bands $n=0-6$ for various well depths. The values for the case of a linearized Boltzmann equation with classical statistics are nearly identical. This table shows the evolution of the slowest eigenvector as various scattering rates change with lattice depth. For purposes of comparison, the normalization condition $\sum_nv_{1n}^2=1$ and a consistent sign convention were used. Note that $v_{11}$ passes through zero slightly above a lattice depth of 10$E_R$.}
\label{tab:eigvector_bose}
\end{table}

\section{\label{sec:conclusions}Conclusions}

We have considered the preparation, short-time evolution and relaxation of the $p$-orbital insulator state. 
A $\pi$-pulse with characteristic time $\tau$ sufficiently long that $\tilde{U}\tau\gg1$ will end in state which is a close approximation to the $p$-orbital insulator, with a density of excitations above the ground state which is exponentially small for large $\tau$. At short times, the insulator state is metastable due to the anharmonicity of the optical lattice. For well depths below the two-body threshold value $V_t$, our theory for the short time decay rate is broadly in agreement with measurements performed by M\"uller et al. \cite{Muller2007}; above threshold there is a marked disagreement, perhaps due to a small fraction of double occupied sites from an imperfect $\pi$-pulse in experiments. Reducing this fraction could lead to substantially longer excited band lifetimes. As discussed in section \ref{sec:loadrelax}, a single-tone Gaussian $\pi$-pulse may provide better fidelity than current experimental pulse shapes. 

The interband transitions lead to evolution of the excited band occupation. This in turn affects the transition rates. The intermediate regime of relaxation is further complicated by a spreading in the magnetic trap, and the dynamics of this intermediate regime remains beyond our theory. At long times the system relaxes to an equilibrium gas of nearly free bosons, where the chemical potential and temperature describing the gas are determined by energy and number conservation. We modeled this asymptotic relaxation using a Boltzmann formalism with both classical and Bose statistics. The calculated decay rates differ considerably from experimental values; this disagreement could come from neglecting spatial variations in the trap.

The double-well optical lattice could offer an improvement for excited band lifetimes in the insulating regime. The double well lattice can satisfy the hard anharmonicity condition, $\epsilon_2+\epsilon_0-2\epsilon_1>0$, for suitable lattice parameters. This condition changes the initial decay mechanism from an exothermic process to an endothermic one, leading to a significant enhancement of the excited band lifetime. Stojanovi\'c et al.'s work in the superfluid regime found that excited band lifetimes in the double-well lattice could be orders of magnitude larger than equivalent single-well lattices \cite{Stojanovic2008}. 

Our theory provides a framework to understand a wide range of experiments performed in the excited band, including those performed by M\"uller et al. \cite{Muller2007}.  We hope these insights and calculations can improve the experimental realization of excited band models in optical lattices. 

\acknowledgments{The authors would like to thank T. M\"uller, S. F\"olling, A. Widera and I. Bloch for graciously sharing their data on excited band lifetime measurements. Additionally, we would like to thank A. Polkovnikov and W. Zwerger for their useful discussions. Finally we wish to acknowledge Victoria Caldwell for her help in preparing the figures. JHC and SMG were supported by NSF DMR-0603369, and LIG was supported by NSF grant DMR-0754613.}
\appendix

\section{Schr\"odinger equation of $\pi$-pulse in two-site model}
\label{sec:appendix_pulse}
As discussed in Section \ref{sec:loadrelax}, we can understand many aspects of the preparation of the \textit{p}-orbital insulator using a model of two bosons in a well of two sites.  Each site has an $s$ level and a $p$ level, but we only include tunneling in the $p$-band. We initially prepare the system in the two-particle ground state, and then promote both particles to the excited level through a $\pi$-pulse. The model Hamiltonian $H$ is given by :
\begin{align}
H=&\epsilon_{10}b_{L1}^{\dag}b_{L1}+t_{1}b_{L1}^{\dag}b_{R1}+\frac{U_{11}}{2}b_{L1}^{\dag}b_{L1}^{\dag}b_{L1}b_{L1}\nonumber\\
&+\frac{U_{00}}{2}b_{L0}^{\dag}b_{L0}^{\dag}b_{L0}b_{L0}+U_{10}b_{L1}^{\dag}b_{L0}^{\dag}b_{L1}b_{L0}\nonumber\\
&+\left(L\rightarrow R\right).
\end{align}
The six relevant parity-symmetric eigenstates are given to $O(t_1/U)$ in Eq.~(\ref{eq:symm_eigenstates}). We promote the bosons from the $s$-band to the $p$-band using a pulse of the form:
\begin{equation}
H_{\textrm{pulse}}=\Delta(t)\ e^{-i\omega t}\left(b_{L1}^{\dag}b_{L0}+b_{R1}^{\dag}b_{R0}\right)+\textrm{h.c.},
\end{equation}
with the Gaussian envelope $\Delta(t)=\frac{\alpha}{\tau}\exp(-t^2/\tau^2)$. We choose the dimensionless pulse strength $\alpha$ to optimize a single $\pi$-pulse . The excitation can be considered a standard two-photon process, with the frequency $\omega$ chosen to be half the energy difference of states $\left|1\right\rangle$ and $\left|5\right\rangle$.

The Hilbert space naturally divides between a subspace composed of states with nearly one particle per site, $\lbrace \left\vert1\right\rangle,\left\vert3\right\rangle,\left\vert5\right\rangle\rbrace$ and a subspace composed of states which include doubly occupied sites $\lbrace \left\vert2\right\rangle,\left\vert4\right\rangle,\left\vert6\right\rangle\rbrace$. To understand the behavior of the Schr\"odinger equation, we first solve for the evolution within the $\lbrace \left\vert1\right\rangle,\left\vert3\right\rangle,\left\vert5\right\rangle\rbrace$ subspace and then use that time evolution as a drive term for the $\lbrace \left\vert2\right\rangle,\left\vert4\right\rangle,\left\vert6\right\rangle\rbrace$ subspace. 

\subsection{Evolution of $\lbrace \left\vert1\right\rangle,\left\vert3\right\rangle,\left\vert5\right\rangle\rbrace$ subspace}
For the following we denote the amplitudes of the $\left\vert1\right\rangle,\left\vert3\right\rangle,\left\vert5\right\rangle\rbrace$ by $\mathbf{a}=\lbrace a_{1}, a_{3}, a_{5} \rbrace$. The Schr\"odinger equation in the $\tilde{U}\rightarrow\infty$ limit becomes
\begin{equation}
 \label{eq:subspace1_schrod}
i\frac{d}{du}\mathbf{a}=H_{135}\mathbf{a},
\end{equation}
with Hamiltonian $H_{135}$ given in the frame rotating at frequency $\omega$ by:
\begin{equation}
H_{135}=\left(\begin{array}{ccc}
0 & \sqrt{2}\alpha e^{-u^{2}} & 0\\
\sqrt{2}\alpha e^{-u^{2}} & \kappa & \sqrt{2}\alpha e^{-u^{2}}\\
0 & \sqrt{2}\alpha e^{-u^{2}} & 0\end{array}\right).
\end{equation}
We operate in the regime $\kappa\ll1$, where we can expand the solution $\mathbf{a}$ as a power series in $\kappa$; see Eq.~(\ref{eq:dimensionless_variables}) for a definition. Changing variables from $u$ to $v$:
\begin{equation}
\label{eq:v_in_u}
v=\alpha\sqrt{\frac{\pi}{2}}\left(1+\textrm{erf}\ u \right),
\end{equation}
where $\textrm{erf}$ denotes the error function. Here $v$ ranges between 0 and $\alpha \sqrt{2\pi}$. In terms of these new variables, Eq.~(\ref{eq:subspace1_schrod}) becomes
\begin{equation}
i\frac{d}{dv}\mathbf{a}=
\left[\left(\begin{array}{ccc}
0 & 1 & 0\\
1 & 0 & 1\\
0 & 1 & 0\end{array}\right)+\left(\begin{array}{ccc}
0 & 0 & 0\\
0 & \frac{\kappa}{\sqrt{2\alpha}}e^{u^2(v)} & 0\\
0 & 0 & 0\end{array}\right)\right]
\mathbf{a},
\label{eq:perturb_de}
\end{equation}
where $u$ is an implicit function of $v$. Since Eq.~(\ref{eq:perturb_de}) obeys the constraint $a_1-a_5=1$, we can simplify the Schr\"odinger equation in terms of the new variable $\mathbf{z}$:
\begin{subequations}
\begin{align}
 \mathbf{z}=&\left(
\begin{array}{c}
             \frac{a_1+a_5-\sqrt{2}a_3}{2}\\
\frac{a_1+a_5+\sqrt{2}a_3}{2}
            \end{array}\right),\\
i\frac{d\mathbf{z}}{dv}=&\left(\begin{array}{cc} 
-\sqrt{2}&0\\
0&\sqrt{2}\end{array}\right) \mathbf{z}+\kappa\frac{e^{u^2(v)}}{2\sqrt{2}\alpha}\left(\begin{array}{cc} 
1&-1\\
-1&1\end{array}\right).
\end{align}
\end{subequations}
Expanding $\mathbf{z}$ as a power series in $\kappa$, we find both the first-order corrections $f_1$ and $f_2$ as well as the second-order corrections $h_1$ and $h_2$:
\begin{subequations}
\begin{align}
 \mathbf{z}=&\left(
\begin{array}{c}
 e^{i\sqrt{2}v}/\sqrt{2}+\kappa f_{1}(v)+\kappa^2h_{1}(v)+\ldots\\
e^{-i\sqrt{2}v}/\sqrt{2}+\kappa f_{2}(v)+\kappa^2h_{2}(v)+\ldots
\end{array}
\right),\\
f_{1}(v)=&\frac{e^{i\sqrt{2}v}}{2\sqrt{2}\alpha}\int_{0}^{v}dv'e^{u^{2}(v')-i\sqrt{2}v'}\sin(\sqrt{2}v'),\\ 
f_{2}(v)=&-f_1^*(v),\\
h_{1}(v)=&\frac{-i}{4\alpha^{2}}\int_{0}^{v}dv'e^{u^{2}(v')-i(v'-v)\sqrt{2}}\int_{0}^{v'}dv''\times\nonumber\\
&e^{u^{2}(v'')}\cos\left[\sqrt{2}(v'-v'')\right]\sin\left(\sqrt{2}v''\right),\\
h_{2}(v)=&h_1^*(v).
\end{align}
\end{subequations}
In terms of the functions $f_1(v)$ and $h_1(v)$, we can express $\lbrace a_1,a_3,a_5\rbrace$ to $O(\kappa^2)$:
\begin{subequations}
\label{eq:adiabatic_de_soln}
\begin{align}
 a_{1}(v)\approx&\cos^2\left(\frac{v}{\sqrt{2}}\right)+i\textrm{Im}[f_1(v)] \kappa+\textrm{Re}[h_1(v)]\kappa^2,\\
a_{3}(v)\approx&-\frac{i\sin(v\sqrt{2})}{\sqrt{2}}-\sqrt{2}\textrm{Re}[f_1(v)]\kappa\nonumber\\
&-i\sqrt{2} \textrm{Im}[h_1(v)] \kappa^2,\\
a_{5}(v)\approx&-\sin^2\left(\frac{v}{\sqrt{2}}\right)+i\textrm{Im}[f_1(v)] \kappa+\textrm{Re}[h_1(v)]\kappa^2.
\end{align}
\end{subequations}
\subsection{Calculating $\left|a_5\right|^2$ after application of pulse} 
We now determine the fraction of particles excited to state $\left\vert5\right\rangle$, the analogue of the $p$-orbital insulator, in the limit $\kappa\ll1$. We use the transformed variable $v$ defined in Eq.~(\ref{eq:v_in_u}). For $\kappa=0$, $\left|a_{5}(\alpha\sqrt{2\pi})\right|^2$ reaches its maximum value of unity when $\alpha=\frac{\sqrt{\pi}}{2}$. Expanding $\alpha$ as a Taylor series in $\kappa$ gives
\begin{equation}
\label{eq:alpha_expans}
\alpha=\frac{\sqrt{\pi}}{2}+b\kappa+c\kappa^2+\ldots\ .
\end{equation}
In terms of these coefficients, the leading order correction to $\left|a_{5}\right|^2$ after the pulse is
\begin{equation}
\left|a_{5}(\alpha\sqrt{2\pi})\right|^2=\left\vert a_5\left(\frac{\pi}{\sqrt{2}}\right)\right\vert^2-2\pi b^2 \kappa^2+O(\kappa^3). 
\end{equation}
To $O(\kappa^2)$, $\left|a_{5}(\alpha\sqrt{2\pi})\right|^2$ is maximized when $b=0$. Our final expression for the occupation $\left|a_5\right|^2$ after the pulse is
\begin{align}
\label{eq:a5_expand}
\left|a_5\left(\frac{\pi}{\sqrt{2}}\right)\right|^2&=1-\left|a_1\left(\frac{\pi}{\sqrt{2}}\right)\right|^2-\left|a_3\left(\frac{\pi}{\sqrt{2}}\right)\right|^2,\\
&=1-(0.265\ldots)\kappa^2+O\left(\kappa^4\right).
\end{align}
\subsection{Calculating the dimensionless pulse strength $\alpha$}
In order to determine the dimensionless pulse strength $\alpha$ to $O(\kappa^2)$, we need to maximize $\left|a_{5}(\alpha \sqrt{2 \pi})\right|^2$ over the coefficient $c$ in Eq.~(\ref{eq:alpha_expans}):
\begin{align}
 \left|a_{5}(\alpha \sqrt{2 \pi})\right|^2=&1-(0.265\ldots)\kappa^2\nonumber\\
&-\kappa^4\left[\left(c\sqrt{2\pi}+\beta\right)^2-\gamma c\right]+O\left(\kappa^6\right),
\end{align}
where the values $\beta$ and $\gamma$ are given by:
\begin{subequations}
\label{eq:beta_gamma}
\begin{align}
\beta&=0.0825\ldots\ ,\\
\gamma&=0.9476\ldots\ .
\end{align}
\end{subequations}
To $O(\kappa^4)$, $\left|a_{5}(\alpha \sqrt{2 \pi})\right|^2$ is maximized when:
\begin{equation}
c=\frac{\gamma-2\sqrt{2\pi}\beta}{4\pi}=0.0425\ldots\ .
\end{equation}
The optimal value of the dimensionless strength $\alpha$ which maximizes the occupation of state $\left|5\right\rangle$ is given to $O(\kappa^2)$ by:
\begin{equation}
\alpha=\frac{\sqrt{\pi}}{2}-(0.0425\ldots)\kappa^2+O\left(\kappa^4\right),
\end{equation}
The corresponding value of $|a_5|^2$ is
\begin{equation}
\left|a_{5}(\alpha \sqrt{2 \pi})\right|^2=1-(0.265\ldots)\kappa^2+(0.0045\ldots)\kappa^4+O(\kappa^6).
\end{equation}

\subsection{Evolution of $\lbrace \left\vert2\right\rangle,\left\vert4\right\rangle,\left\vert6\right\rangle\rbrace$ subspace}
\label{sec:scratchwork_subspace_two}
\begin{figure}
 \centering
 \includegraphics[width=1.00\columnwidth]{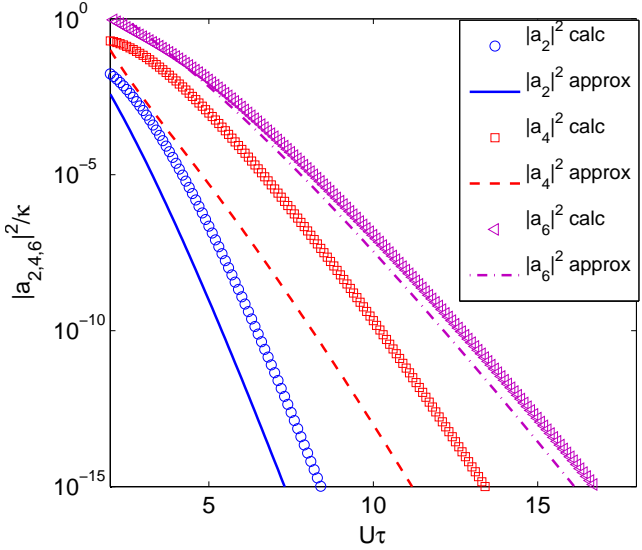}
 \caption{(Color online.) Comparison of the numerical calculation of $\lim_{\kappa\rightarrow0}|a_{n}(\infty)|^2/\kappa$ for $n=2,4,6$ with the stationary phase approximation; see Eqs.~(\ref{eq:a2_a4_a6_full}) and (\ref{eq:a2_a4_a6_statphase}).}
 \label{fig:a2_a4_a6_abs}
\end{figure}

To estimate the amplitudes $a_2$, $a_4$, and $a_6$ of the $\lbrace \left\vert2\right\rangle,\left\vert4\right\rangle,\left\vert6\right\rangle\rbrace$ subspace, we substitute the leading terms for amplitudes $\lbrace a_1,a_3,a_5\rbrace$ from Eq.~(\ref{eq:adiabatic_de_soln}) into the Schr\"odinger equation. This procedure is valid to leading order in the small parameter $\kappa/U$:
\begin{subequations}
\label{eq:a2_approx}
\begin{align}
 \frac{da_2}{du}&\approx -iU_{2}a_2+\frac{\sqrt{\pi}t_1}{2\sqrt{2}U_{10}}e^{-u^2}\cos\left(\frac{\pi}{2}\textrm{erf }u\right),\\
\frac{da_4}{du}&\approx-i\frac{\sqrt{\pi}t_1}{\sqrt{2}U_{10}}e^{-u^2}\cos^2\left[\frac{\pi}{4}\left(1+\textrm{erf }u\right)\right]-iU_4a_4\nonumber\\
&-i\frac{\sqrt{2\pi}t_1}{U_{11}}\left(1-\frac{U_{11}}{2U_{10}}\right)e^{-u^2}\sin^2\left[\frac{\pi}{4}\left(1+\textrm{erf }u\right)\right],\\
\frac{da_6}{du}&\approx-\frac{\sqrt{2\pi}t_1}{U_{11}}\left(1-\frac{U_{11}}{4U_{10}}\right)e^{-u^2} \cos\left(\frac{\pi}{2}\textrm{erf } u\right)-iU_6a_6,
\end{align}
\end{subequations}
with $\lbrace U_2,U_4,U_6\rbrace$ defined by
\begin{subequations}
\begin{align}
U_{2}=&\frac{U_{00}}{U_{11}}\tilde{U},\\
U_{4}=&\frac{U_{10}}{U_{11}}\tilde{U}+\kappa\left(\frac{2\frac{U_{10}}{U_{11}}+1}{2\frac{U_{10}}{U_{11}}-1}\right),\\
U_6=&\tilde{U}+\frac{8\kappa}{2-\frac{U_{11}}{U_{10}}}.
\end{align}
\end{subequations}
From Eq.~(\ref{eq:dimensionless_variables}), we make the connection between $t_1/U_{11}$ and $\kappa/\tilde{U}$ :
\begin{equation}
 \frac{t_1}{U_{11}}=\sqrt{\frac{\kappa}{\tilde{U}\left(2-\frac{U_{11}}{U_{10}}\right)}}.
\end{equation}
We expand $\lbrace a_2,a_4,a_6\rbrace$ to leading order in $\kappa/U$:
\begin{equation}
\begin{array}{cc}
\left|a_n(\infty)\right|^2=\frac{\kappa}{\tilde{U}}\left|a_n^{(1)}\right|^2+O\left[\left(\kappa/\tilde{U}\right)^{2}\right] &\text{if }n=2,4,6\ .
\end{array}
\end{equation}
Solving Eq.~(\ref{eq:a2_approx}) gives us the expressions for $\lbrace a_{2}^{(1)},a_{4}^{(1)},a_{6}^{(1)}\rbrace$:
\begin{subequations}
\label{eq:a2_a4_a6_full}
\begin{align}
 \left|a_2^{(1)}\right|^2&=\frac{\pi r^2}{8\left(2-r\right)}\left|\int_{-\infty}^{\infty}due^{-u^2+iU_2u}\cos\left(\frac{\pi \textrm{erf }u}{2}\right)\right|^2,\\
\left|a_4^{(1)}\right|^2&=\frac{2\pi}{2-r}\left|\int_{-\infty}^{\infty}du\left\{\frac{r}{2}\cos^2\left[\frac{\pi\left(1+\textrm{erf }u\right)}{4}\right]\right.\right.\nonumber\\
&\left.\left.+\left(1-\frac{r}{2}\right)\sin^2\left[\frac{\pi\left(1+\textrm{erf }u\right)}{4}\right]\right\}e^{-u^2+iU_4u}\right|^2,\\
\left|a_6^{(1)}\right|^2&=\frac{\pi \left(1-\frac{r}{4}\right)^2}{\left(1-\frac{r}{2}\right)}\left|\int_{-\infty}^{\infty}due^{-u^2+iU_6u}\cos\left(\frac{\pi \textrm{erf }u}{2}\right)\right|^2,
 \end{align}
\end{subequations}
with $r=U_{11}/U_{10}$. In the limit of large $\tilde{U}$, $a_2^{(1)}$,$a_4^{(1)}$ and $a_6^{(1)}$ are all exponentially suppressed. We can approximate these terms using the stationary phase approximation (see Appendix \ref{sec:stat_phase}): 
\begin{widetext}
\begin{subequations}
\label{eq:a2_a4_a6_statphase}
 \begin{align}
\left|a_2(\infty)\right|^2\approx&\frac{\pi \kappa U_{11}^2}{8\tilde{U}\left(2U_{10}^2-U_{11}U_{10}\right)}\frac{U_2\exp\left[-2U_2x_0(U_2)\right]}{x_0(U_2)}\cosh^2\left\{\frac{U_2}{2x_0(U_2)}\left[1+\frac{1}{2x_0(U_2)^2}\right]\right\},\\
\left|a_4(\infty)\right|^2\approx&\frac{\pi \kappa U_{10}}{2\tilde{U}\left(2U_{10}-U_{11}\right)}\frac{U_4\exp\left[-2U_4x_0(U_4)\right]}{x_0(U_4)}\left(1+\left(1-\frac{U_{11}}{U_{10}}\right)^2\sinh^2\left\{\frac{U_4}{2x_0(U_4)}\left[1+\frac{1}{2x_0(U_4)^2}\right]\right\}\right),\\
\left|a_6(\infty)\right|^2\approx&\frac{2\pi \kappa U_{10}}{\tilde{U}\left(2U_{10}-U_{11}\right)}\left(1-\frac{U_{11}}{4U_{10}}\right)^2\frac{U_6\exp\left[-2U_6x_0(U_6)\right]}{x_0(U_6)}\cosh^2\left[\frac{\pi}{2}\textrm{erfi }x_0(U_6)\right],
\end{align}
\end{subequations}
\end{widetext}
with the function $x_0(U)$ given by:
\begin{equation}
 x_0(U_n)=\sqrt{\log\frac{U_n}{\sqrt{\pi}}}.
\end{equation}
In Fig. \ref{fig:a2_a4_a6_abs} we compare numerical determinations of $\lim_{\kappa\rightarrow0}\left|a_n\right|^2/\kappa$ with a stationary phase approximation calculated in Eq.~(\ref{eq:a2_a4_a6_statphase}) Since $U_6<U_2,U_4$, $a_6\gg a_2,a_4$ due to the exponential dependence on $U_n$. This makes $a_6^{(1)}$ the most important error in determining $\left|a_5(\infty)\right|^2$:
\begin{align}
 \left|a_5(\infty)\right|^2&=1-\frac{\kappa}{\tilde{U}}\left(\left|a_2^{(1)}\right|^2+\left|a_4^{(1)}\right|^2+\left|a_6^{(1)}\right|^2\right)\nonumber\\
&-(0.265\ldots)\kappa^2+(0.0045\ldots)\kappa^4+O\left(\kappa^6,\frac{\kappa^2}{\tilde{U}^2}\right).
\end{align}

\section{Stationary Phase Integration}
\label{sec:stat_phase}
The following integral frequently arises in our discussion of the preparation of the $p$-orbital insulator using $\pi$-pulses with Gaussian envelopes:
\begin{equation}
\label{eq:stat_phase}
Y(k)=\int_{-\infty}^{\infty}dx \exp\left[-x^2+ikx-i\frac{\pi}{2}\textrm{erf}(x)\right].
\end{equation}
We want to derive an asymptotic expansion for $Y(k)$ in the limit $k\gg1$. We define the inverse function for $y(x)=e^{-x^2}$ at all points in the complex plane by choosing the branch cut to be along the negative imaginary axis:
\begin{equation}
y^{-1}(x)=i\sqrt{\log x}.
\end{equation}
$Y(k)$ has a stationary phase point at $x_{1}$:
\begin{equation} 
x_{1}=i\left[\sqrt{\log\frac{k}{\sqrt{\pi}}}-\frac{1}{k}+O\left(\frac{\sqrt{\log k}}{k^2}\right)\right].
\end{equation}
This formula is valid for $k$ both positive and negative. Using the standard stationary phase approximation, we can write an expansion for asymptotically large $k$: 
\begin{align}
\label{eq:y_neg}
Y(k)\approx\frac{\sqrt{k}}{(\log\frac{k}{\sqrt{\pi}})^{1/4}}\exp\left[-\vert k\vert\sqrt{\log\frac{k}{\sqrt{\pi}}}\right.\times\nonumber\\
\left.\left(1-\frac{1}{2\log \frac{k}{\sqrt{\pi}}}-\frac{1}{4\log^2\frac{k}{\sqrt{\pi}}}+\ldots\right)\right].
\end{align}
From Eq.~(\ref{eq:y_neg}), we see $Y(k)$ decreases faster for large negative $k$ than large positive $k$ because the magnitude of the logarithm term is larger:
\begin{equation}
 \left\vert\log\frac{-|k|}{\sqrt{\pi}}\right\vert=\sqrt{\log^2\frac{|k|}{\sqrt{\pi}}+\pi^2}.
\end{equation}
A more qualitative way to understand this result is to expand the exponential of $Y(k)$'s integrand in Eq.~(\ref{eq:stat_phase}) for small $x$:
\begin{equation}
 -x^2+ikx-i\frac{\pi}{2}\textrm{erf}(x)\approx-x^2+i(k-\sqrt{\pi})x.
\end{equation}
The integrand is more slowly oscillating for $k>0$ than for $k<0$. 

\subsection{Related Integrals}
In the course of computing the density of excitations in section \ref{sec:full_adiabatics}, the following two integrals $c(k)$ and $s(k)$ arise:
\begin{subequations}
\begin{align}
c(k)&=\int_{-\infty}^{\infty}due^{iku-u^2}\cos\left(\frac{\pi}{2}\textrm{erf }u\right),\\
&=\frac{Y(k)+Y(-k)^*}{2},\\
s(k)&=\int_{-\infty}^{\infty}due^{iku-u^2}\sin\left(\frac{\pi}{2}\textrm{erf }u\right),\\
&=\frac{-Y(k)+Y(-k)^*}{2i}.
\end{align}
\end{subequations}
Both integrals share the same stationary phase point $x_1$:
\begin{equation}
x_{1}=i\sqrt{\log\frac{k}{\sqrt{\pi}}}+O\left(\frac{1}{k}\right).
\end{equation}
In the stationary phase approximation, the integrals become
\begin{subequations}
\label{eqnarray:cosh_int}
\begin{align}
c(k)=&\sqrt{\frac{k}{\sqrt{\log\frac{k}{\sqrt{\pi}}}}}\exp\left(-k\sqrt{\log\frac{k}{\sqrt{\pi}}}\right)\times\nonumber\\
&\cosh\left[\frac{k}{2\log\frac{k}{\sqrt{\pi}}}\left(1+\frac{1}{2\log\frac{k}{\sqrt{\pi}}}+\ldots\right)\right],\\
s(k)=&i\sqrt{\frac{k}{\sqrt{\log\frac{k}{\sqrt{\pi}}}}}\exp\left(-k\sqrt{\log\frac{k}{\sqrt{\pi}}}\right)\times\nonumber\\
&\sinh\left[\frac{k}{2\log\frac{k}{\sqrt{\pi}}}\left(1+\frac{1}{2\log\frac{k}{\sqrt{\pi}}}+\ldots\right)\right].
\end{align}
\end{subequations}
\section{Deriving recursion relations for $P_g^n$ and $\rho_{\textrm{exc}}$}
\label{sec:diff_eqns_appendix}
In Section \ref{sec:full_adiabatics}, we consider the problem of loading the excited band in an system of $N$ sites. Naive perturbation techniques fail because the expansion parameter changes from $t_1/U$ to $t_1\sqrt{N}/U$, a large parameter in the thermodynamic limit. We get around this difficulty by first allowing hopping only between $n$ sites, where $n$ is sufficiently small that $t_1\sqrt{n}/U\ll1$ and perturbation theory remains valid. We derive recursion relations in the variable $n$ for two important quantities: the probability $P_g$ that the final state after application of the pulse is the $p$-orbital insulator and the ``density of excitations'' $\rho_{\textrm{exc}}$ of the final state defined in Eq.~(\ref{eq:rho_exc_def}). By solving these relations for $n\rightarrow N$, we derive expressions for the probability $P_g$ and the density of excitations $\rho_{\textrm{exc}}$ accurate in the thermodynamic limit. 

An excitation pulse $\Delta(t)e^{i\omega t}$ promotes bosons from the ground band to the first excited band (see Fig. \ref{fig:nsite_adiabatics}). We choose the pulse carrier frequency $\omega$ to maximize the occupation of the $p$-orbital insulator. We are interested in pulses sufficiently long that $\tilde{U}=U\tau\gg1$, so that the excitation density above the ground state is small. In the following discussion, the phase angle $\theta(t)=\int_{-\infty}^tdt'\Delta(t')$ frequently arises. 

\begin{figure}
 \centering
 \includegraphics[width=1.00 \columnwidth]{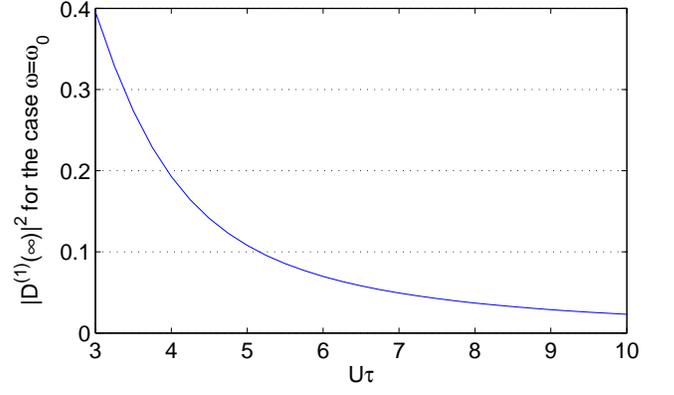}
\caption{The coefficient $\left\vert D^{(1)}(\infty)\right\vert^2$ for the specific case of a Gaussian pulse; see Eq.~(\ref{eq:d_1_expression}). As expected, it is strongly suppressed for large $U\tau$.}
\label{fig:d1_abs_sq}
\end{figure}

By changing to the interaction picture, we transform the time dependence of the pulse into the time dependence of the tunneling $t_1$ and the pulse detuning $\epsilon_{10}-\omega$. Treating all of the Hubbard $U$ parameters as equal, the Hamiltonian $H_n$ with tunneling only allowed between the first $n$ nearest neighbor pairs [see Eq.~(\ref{eq:hamiltonian_n})] is given by:
\begin{align}
\label{eq:ham_near_neighbor_appendx}
H_n&(t)=t_1\sum_{j=1}^n\left(c b^{\dag}_{1j+1}+is\ b^{\dag}_{0j+1}\right)\left(c\ b_{1j}-is\ b_{0j}\right)+\textrm{h.c.}\nonumber\\
&+\left(\epsilon_{10}-\omega\right)\sum_{j=1}^n\left(c\ b^{\dag}_{1j}+is\ b^{\dag}_{0j}\right)\left(c\ b_{1j}-is\ b_{0j}\right)+\textrm{h.c.}\nonumber\\
&+\frac{U}{2}\sum_j\sum_{nn'}b_{nj}^{\dag}b_{n'j}^{\dag}b_{n'j}b_{nj},
\end{align}
where $c=\cos\theta$ and $s=\sin\theta$. Initially, we prepare the system in the true $N$-particle ground state $\left|\psi_0\right\rangle$ at a density of one particle per site:
\begin{equation}
\left|\psi_0\right\rangle=\prod_{j=1}^Nb_{0j}^{\dag}\left\vert0\right\rangle.
\end{equation}
The wave function $\left\vert\Psi_I^n\right\rangle$ describes the evolution of the system under $H_n$.
Expanding $\left\vert\Psi_I^n\right\rangle$ to leading order in the small parameters $t_1\sqrt{n}/U$ and $t_1^2\tau\sqrt{n}/U$ gives
\begin{align}
\label{eq:leading_order_n_pairs}
\left|\Psi_I^n(t)\right\rangle=&A_n(t)\prod_{j'=1}^Nb_{0j'}^{\dag}\left\vert0\right\rangle\nonumber\\
&+\frac{t_1}{U}B^{(1)}(t)\sum_{j=1}^n\left(b_{1j+1}^{\dag}+b_{1j-1}^{\dag}\right)b_{0j}\prod_{j'=1}^Nb_{0j'}^{\dag}\left\vert0\right\rangle\nonumber\\
&+\frac{t_1}{U}C^{(1)}(t)\sum_{j=1}^n\left(b_{0j+1}^\dag+b_{0j-1}^\dag\right)b_{0j}\prod_{j'=1}^Nb_{0j'}^{\dag}\left\vert0\right\rangle\nonumber\\
&+\frac{t_1^2\tau}{U}D^{(1)}(t)\sum_{j=1}^n b_{1j}^{\dag}b_{0j}\prod_{j'=1}^Nb_{0j'}^{\dag}\left\vert0\right\rangle\nonumber\\
&+O\left(\frac{nt_1^2}{U^2},\frac{nt_1^4\tau}{U^2}\right).
\end{align}
The coefficients $B^{(1)}$, $C^{(1)}$ and $D^{(1)}$ must all be independent of $n$ by translational invariance, whereas $A_n$ must decrease with increasing $n$ in order to preserve normalization. Expanding the Schr\"odinger equation for $B^{(1)}$, $C^{(1)}$, and $D^{(1)}$ to leading order in small parameters gives
\begin{subequations}
\label{eq:B_adiabatic}
\begin{align}
\label{eq:B_adiabatic_b}
 i\frac{d}{du}B^{(1)}&\approx \tilde{U}\left(i\sqrt{2}\sin\theta\cos\theta+B^{(1)}\right),\\
\label{eq:B_adiabatic_c}
i\frac{d}{du}C^{(1)}&\approx \tilde{U}\left(2\sin^2\theta+C^{(1)}\right),\\
i\frac{d}{du}D^{(1)}&\approx -i\left(\frac{\omega-\epsilon_{10}}{t_1^2/U}\right)\sin\theta\cos\theta+\sqrt{2}\cos^2\theta B^{(1)}\nonumber\\
&+i\sin\theta\cos\theta C^{(1)}-i\left(\frac{\omega-\epsilon_{10}}{t_1^2/U}\right)\cos^2\theta D^{(1)}.
\end{align}
\end{subequations}
Multiplying both sides of Eqs.~(\ref{eq:B_adiabatic_b}) and (\ref{eq:B_adiabatic_c}) by $e^{i\tilde{U}u}$ and integrating by parts gives
\begin{subequations}
\label{eq:approx_de}
 \begin{align}
  B^{(1)}=&-i\sqrt{2}\sin\theta\cos\theta+\delta B^{(1)},\\
C^{(1)}=&-2\sin^2\theta+\delta C^{(1)},
 \end{align}
\end{subequations}
where the terms $\delta B^{(1)}$ and $\delta C^{(1)}$ are given by:
\begin{subequations}
 \begin{align}
\delta B^{(1)}=&i\sqrt{2}e^{-i\tilde{U}u}\int_{-\infty}^udu'\dot{\theta}\cos2\theta e^{i\tilde{U}u'},\\
\delta C^{(1)}=&2e^{-i\tilde{U}u}\int_{-\infty}^udu'\dot{\theta}\sin2\theta e^{i\tilde{U}u'}.
 \end{align}
\end{subequations}
The terms $\delta B^{(1)}$ and $\delta C^{(1)}$ in Eq.~(\ref{eq:approx_de}) represent excitations above the ground state and so contribute to the density of excitations $\rho_{\textrm{exc}}$. Substituting our solutions for $B^{(1)}$ and $C^{(1)}$ from Eq.~(\ref{eq:approx_de}) into Eq.~(\ref{eq:B_adiabatic}), we get the following expression for $D^{(1)}(\infty)$:
\begin{subequations}
\label{eq:d_1_expression}
\begin{align}
D^{(1)}(\infty)=&-\left[\left(\frac{\omega-\epsilon_{10}}{t_1^2/U}\right)+2\right]I_1+I_2,\\
I_1=&\int_{-\infty}^{\infty}du\sin\theta(u)\cos\theta(u),\\
I_2=&\int_{-\infty}^{\infty}du\int_{-\infty}^{u}du'e^{i\tilde{U}(u'-u)}\dot{\theta}(u')\nonumber\times\\
&\left[2\cos^2\theta (u)\cos 2\theta(u')+\sin2\theta(u)\sin2\theta(u')\right].
\end{align}
\end{subequations}
To minimize $\left|D^{(1)}(\infty)\right|^2$, we choose the frequency $\omega$ to cancel the real part of $D^{(1)}(\infty)$:
\begin{equation}
\label{eq:frequency_shift}
 \omega=\epsilon_{10}-\frac{t_1^2\tau}{U}\left(2-\frac{\textrm{Re}\left[I_2\right]}{I_1}\right).
\end{equation}
For $\tilde{U}\gg1$, Eq.~(\ref{eq:frequency_shift}) reproduces the two-site result of $\omega\approx\epsilon_{10}-2t_1^2\tau/U$. For this optimal frequency, the value of $D^{(1)}(\infty)$ becomes
\begin{subequations}
\begin{align}
\left|D^{(1)}(\infty)\right|^2&=\left|\int_{-\infty}^{\infty}du\int_{-\infty}^{u}du'\sin\left[\tilde{U}(u'-u)\right]\dot{\theta}(u')\nonumber\times\right.\\
&\left\{\cos2\theta(u')+\cos\left[2\theta(u)-2\theta(u')\right]\right\}\bigg|^2.
\end{align}
\end{subequations}
We have plotted $\left\vert D^{(1)}(\infty)\right\vert^2$ for the specific case of a Gaussian pulse in Fig. \ref{fig:d1_abs_sq}.

We can determine the ground state wave function $\left|\Psi_g^n\right\rangle$ of $H_n(\infty)$ by enforcing the adiabatic pulse condition: 
\begin{subequations}
\begin{align}
\left|\Psi_g^n\right\rangle&=\lim_{\tilde{U}\rightarrow\infty}\left\vert\Psi_I^n\right\rangle,\\
&=\prod_{j=1}^Nb_{0j}^{\dag}\left\vert0\right\rangle-\frac{t_1}{U}\sum_{j=1}^n\left(b_{0j+1}^{\dag}b_{0j}+\textrm{h.c.}\right)\prod_{j=1}^Nb_{0j}^{\dag}\left\vert0\right\rangle\nonumber\\&+O\left(\frac{t_1^2}{U^2}\right).
\end{align}
\end{subequations}
In order to compute the probability $P_g$ of completing the pulse in the $p$-orbital insulator state and the density of excitations $\rho_{\textrm{exc}}$, it is convenient to reexpress $ \left|\Psi_I^{n}(\infty)\right\rangle$ from Eq.~(\ref{eq:leading_order_n_pairs}) using $\left|\Psi_g^n\right\rangle$:
\begin{align}
\label{eq:psi_exc_gs}
 \left|\Psi_I^{n}(\infty)\right\rangle&=A_{n}(\infty)\left|\Psi_g^{n}\right\rangle+\frac{t_1^2\tau}{U} D^{(1)}(\infty)\sum_{j=1}^{n}b_{1j}^{\dag}b_{0j}\left|\Psi_g^{n}\right\rangle\nonumber\\
&+\frac{t_1}{U\sqrt{2}}\delta B^{(1)}(\infty)\sum_{j=1}^{n}\left(b_{1j+1}^{\dag}+b_{1j-1}^\dag\right)b_{0j}\left|\Psi_g^{n}\right\rangle\nonumber\\
&+\frac{t_1}{2U}\delta C^{(1)}(\infty)\sum_{j=1}^{n}\left(b_{0j+1}^{\dag}+b_{0j-1}^{\dag}\right)b_{0j}\left|\Psi_g^{n}\right\rangle\nonumber\\
&+O\left(\frac{nt_1^2}{U^2},\frac{nt_1^4\tau^2}{U^2}\right).
\end{align}
To derive the recursion relations, we first calculate the change in $A_n$ as $n\rightarrow n+1$ to preserve normalization:
\begin{align}
\label{eq:amp_diff}
 \frac{A_{n+1}(\infty)}{A_n(\infty)}=1-\frac{t_1^2}{2U^2}\left|\delta B^{(1)}(\infty)\right|^2-\frac{t_1^2}{2U^2}\left|\delta C^{(1)}(\infty)\right|^2\nonumber\\
-\frac{t_1^4\tau^2}{U^2}\left|D^{(1)}(\infty)\right|^2+O\left(\frac{t_1^4}{U^4},\frac{t_1^6\tau^2}{U^4}\right).
\end{align}
Taking the modulus square of Eq.~(\ref{eq:amp_diff}) gives us the recursion relation for $P_g^n$:
\begin{align}
\log \frac{P_g^{n+1}}{P_g^n}=&-\frac{t_1^2}{U^2}\left(\left\vert \delta B^{(1)}(\infty)\right\vert^2+\left\vert \delta C^{(1)}(\infty)\right\vert^2\right)\nonumber\\
&-\frac{2t_1^4\tau^2}{U^2}\left|D^{(1)}(\infty)\right|^2+O\left(\frac{t_1^4}{U^4},\frac{t_1^6\tau^2}{U^4}\right).
\end{align}
To derive the recursion relation for the density of excitations $\rho_{\textrm{exc}}$, we use the definition in Eq.~(\ref{eq:rho_exc_def}) and the wave function $\left\vert\Psi_I^{n+1}\right\rangle$ from Eq.~(\ref{eq:psi_exc_gs}):
\begin{align}
\rho_{\textrm{exc}}(n+1)=&\rho_{\textrm{exc}}(n)+\frac{t_1^2}{NU^2}
\left\vert \delta B^{(1)}(\infty)\right\vert^2\nonumber\\
&+\frac{t_1^2}{NU^2}\left\vert \delta C^{(1)}(\infty)\right\vert^2+O\left(\frac{t_1^4}{NU^4}\right).
\end{align}
Note that we do not include $|D^{(1)}(\infty)|^2$ in the expression above as it is $O(\kappa^2)$, negligible in experimentally interesting regimes. 

\section{Model Demonstrating $R_1\gg R_2$}
\label{sec:r1r2_model}
In section \ref{sec:initstage}, we discussed that the $p$-orbital insulator can directly decay via the $1+1\rightarrow0+2$ channel into two energetically permissible final states, $\left|\psi_{Af}\right\rangle$ and $\left|\psi_{Bf}\right\rangle$ distinguished only by the nature of the holes in the $n=1$ band. One $n=1$ hole is localized around the $n=0$ particle in a bound state in state $\left|\psi_{Af}\right\rangle$, whereas the $n=1$ hole and $n=0$ particle are well separated in state $\left|\psi_{Bf}\right\rangle$. We denote the decay rate to $|\psi_{Af}\rangle$ by $R_1$ and the decay rate to $|\psi_{Bf}\rangle$ by $R_2$. In general, $R_2$ is $O(t_1/U)^2$ smaller than $R_1$, and so can be neglected. We demonstrate this with a simple three-site model that has the essential features of the full N-site system.

Our model consists of three sites at $j=-1,0,1$, each with $n=0$ and $n=1$ band levels. The sites are linked by nearest neighbor tunneling $t_1$ in the excited band. The on-site interaction between particles in bands $n_1$ and $n_2$ is $U_{n_1n_2}$ In addition, we treat the $n=2$ band as a wide continuum, indexed by the quasimomentum $q$. To understand the $1+1\rightarrow0+2$ decay channel, we introduce an interband term $U_{11}^{02}$ that scatters a pair of central site $n=1$ band particles into an $n=2$ band continuum state and a localized $n=0$ band particle. The Hamiltonian neatly divides into two pieces: $H_0$ which preserves band index and $H_1$ which leads to interband transitions. In second quantized notation, $H_0$ and $H_1$ are given by:
\begin{subequations}
\begin{align}
H_0=&\sum_{j=0,\pm1}\epsilon_1b_{1j}^{\dag}b_{1j}+\epsilon_0b_{10}^{\dag}b_{10}+\sum_q\epsilon_{2q}b_{2q}^{\dag}b_{2q}\nonumber\\
&+t_1\left[b_{10}^{\dag}\left(b_{1-1}+b_{11}\right)+\textrm{h.c.}\right]\nonumber\\
&+\frac{U_{11}}{2}b_{10}^{\dag}b_{10}^{\dag}b_{10}b_{10}+U_{10}b_{00}^{\dag}b_{10}^{\dag}b_{00}b_{10},\\
H_1=&U_{11}^{02}\frac{1}{\sqrt{N}}\sum_qb_{2q}^{\dag}b_{0}^{\dag}\left(b_{10}\right)^2+\textrm{h.c.},
\end{align}
\end{subequations}
where the operator $b_{nj}^{\dag}$ creates a particle in band $n$ at site $j$. We initially prepare the system in $\left|\psi_0\right\rangle$, the lowest energy eigenstate of $H_0$ that contains three particles all with band $n=1$ indices. This eigenstate is the model analog of the $p$-orbital insulator. To leading order in $t_1/U$, the eigenstate $\left\vert\psi_0\right\rangle$ and its energy are given by:
\begin{subequations}
\begin{align}
&\left|\psi_0\right\rangle\approx \left[b_{1-1}^{\dag}b_{10}^{\dag}b_{11}^{\dag}-\frac{t_1}{U_{11}}\left(b_{10}^{\dag}\right)^2\left(b_{1-1}^{\dag}+b_{11}^{\dag}\right)\right]\left|0\right\rangle,\\
&\left\langle\psi_0\right|H_0\left|\psi_0\right\rangle\approx3\epsilon_1-\frac{4t_1^2}{U_{11}}.
\end{align}
 \end{subequations}
We denote the $H_0$ eigenstate analogous to $\left|\psi_{Af}\right\rangle$ by $\left|\psi_{2q}\right\rangle$ and the eigenstate analogous to $\left|\psi_{Bf}\right\rangle$ by $\left|\psi_{3q'}\right\rangle$. To leading order in $t_1/U$, these eigenstates and their energies are given by:
\begin{subequations}
\begin{align}
 &\left|\psi_{2q}\right\rangle\approx b_{2q}^{\dag}b_{00}^{\dag}\left[b_{10}\left(\frac{b_{11}+b_{1-1}}{\sqrt{2}}\right)-\frac{t_1\sqrt{2}}{U_{11}}b_{11}b_{1-1}\right]\left|\psi_0\right\rangle,\\
&\left|\psi_{3q'}\right\rangle\approx b_{2q'}^{\dag}b_{00}^{\dag}\left[b_{11}b_{1-1}+\frac{t_1}{U_{11}}b_{10}\left(b_{11}+b_{1-1}\right)\right]\left|\psi_0\right\rangle,\\
&\left\langle\psi_{2q}\right|H_0\left|\psi_{2q}\right\rangle\approx\epsilon_{2q}+\epsilon_1+\epsilon_0-\frac{2t_1^2}{U_{10}},\\
&\left\langle\psi_{3q'}\right|H_0\left|\psi_{3q'}\right\rangle\approx\epsilon_{2q'}+\epsilon_1+\epsilon_0+U_{10}+\frac{2t_1^2}{U_{10}}.
\end{align}
\end{subequations}
The width $4t_2$ of the $n=2$ band is sufficiently large that both $\left|\psi_{2q}\right\rangle$ and $\left|\psi_{3q'}\right\rangle$ conserve energy for specific values of $q,q'$. We use Fermi's Golden Rule to compute the decay rates $R_1$ and $R_2$ from $\left|\psi_0\right\rangle$ to $\left|\psi_{2q}\right\rangle$ and $\left|\psi_{3q'}\right\rangle$ respectively. 
\begin{subequations}
\begin{align}
&\left\langle\psi_{2q}\left|H_1\right|\psi_0\right\rangle\approx\frac{-2t_1}{U_{11}}U_{11}^{02},\\
&\left\langle\psi_{3q'}\left|H_1\right|\psi_0\right\rangle\approx\frac{-2\sqrt{2}t_1^2}{U_{10}U_{11}}U_{11}^{02},\\
&R_1\sim\frac{\left|\left\langle\psi_{2q}\left|H_1\right|\psi_0\right\rangle\right|^2}{\hbar t_2},\\
&R_2\sim\frac{\left|\left\langle\psi_{3q'}\left|H_1\right|\psi_0\right\rangle\right|^2}{\hbar t_2}.
\end{align}
\end{subequations}
Since the matrix element $\left\langle\psi_{3q'}\left|H_{1}\right|\psi_0\right\rangle$ is $O(t_1/U)$ smaller than $\left\langle\psi_{2q}\left|H_1\right|\psi_0\right\rangle$, the rate $R_2$ is $O(t_1/U)^2$ smaller than the rate $R_1$ and can be neglected. This insight holds for the full $N$-site problem considered in Section \ref{sec:initstage}.

\section{Decay rate at short times near two-body threshold $V_t$}
\label{sec:threshold_behavior}
We discuss the behavior of the decay rate at short times for well depths near the two-body threshold value $V_t$. We approximate this decay rate using the two-body rate $1+1\rightarrow0+2$ for well depths below the threshold value $V_t$ and the three-body rate $1+1+1\rightarrow0+0+4$ for well depths above threshold. As the well depth $V\rightarrow V_t^-$ , the two-body rate rises linearly to a constant. However for $V\rightarrow V_t^+$, the three-body rate instead diverges as $[(V-V_t)/t_1]^{-1}$.

\subsection{Two-body decay rate as $V\rightarrow V_t^-$}
\label{subsec:threshold_from_below}
 To evaluate the two-body decay rate near threshold, we use the tight-binding approximation with mean field shifts and neglect hopping in band $n=0$. The band dispersion in the lowest three bands is given by:
\begin{subequations}
\begin{align}
\epsilon_{0k}(V)&\approx\epsilon_0(V),\\
\epsilon_{1k}(V)&\approx\epsilon_{1}(V)+2t_1(V) \cos(\pi k_1),\\
 \tilde{\epsilon}_{2k_2}(V)&\approx
\tilde{\epsilon}_2(V)-2\tilde{t}_2(V)\cos(\pi k_2).
\end{align}
\end{subequations}
The Fermi's Golden Rule formula for the two-body short time decay rate $\Gamma$ is
\begin{align}
 \Gamma=&\frac{32 \pi t_1^2}{\hbar U_{11}^2}\int_{0}^{1}dk_1 \cos^2\pi k_1 \int_{0}^{1} dk_2 \left|U_{11}^{02}(k_2) \right|^2\nonumber\\
&\times\delta\left(\tilde{\epsilon}_2-2\tilde{t}_2\cos\pi k_2+\epsilon_0-2\epsilon_{1}-2t_1\cos\pi k_1\right).
\end{align}
We denote the maximum of the delta function argument as $g(V)$: 
\begin{equation}
\label{eq:g_def}
g(V)=\tilde{\epsilon}_{2}+2\tilde{t}_{2}+\epsilon_{0}-2\epsilon_{1}+2t_{1}.
\end{equation}
If $g(V)\geq 0$, there exists a region in parameter space where the delta function is satisfied and the two-body rate $\Gamma$ is nonzero. For $g(V)<0$, the delta function's argument is never zero and $\Gamma$ vanishes. The condition $g(V_t)=0$ sets the threshold value $V_t$. For the experimental parameters in \cite{Muller2007}, $V_t=25.97 E_R$.

For well depths near threshold, the integrand is nonzero only for $k_2\approx1$; as $U_{11}^{02}(k)$ is relatively insensitive to $k$, we can pull $\left|U_{11}^{02}(k_2=1)\right|^2$ outside the integral. Evaluating the delta function and substituting the variable $y=\sqrt{4t_1/g}\cos \frac{\pi k_1}{2}$ gives:
\begin{equation} \Gamma=\frac{2}{\pi}\Gamma_0\int_{0}^{1}\frac{dy\left(1-\frac{g}{t_1}y^2+\frac{g^2}{4t_1^2}y^4\right)}{\sqrt{\left(1-y^2\right)\left(1-\frac{g}{4t_1}y^2\right)\left(1-\frac{g}{4t_2}+\frac{g}{4t_2}y^2\right)}},
\end{equation}
where the prefactor $\Gamma_0$ is given by:
\begin{equation}
\Gamma_0=\frac{8\left|U_{11}^{02}(1)\right|^2}{\hbar U_{11}^2}\frac{t_1^{3/2}}{\tilde{t}_2^{1/2}}.
\end{equation}
Since $g(V)\ll t_1$ close to threshold, we can expand the two-body decay rate as a power series in $g/t_1$:
\begin{align}
\Gamma=&\frac{2}{\pi}\Gamma_0\int_{0}^{1}\frac{dy}{\sqrt{1-y^2}}\times\nonumber\\
&\left\{1+\frac{g}{t_1}\left[-\frac{7}{8}y^2-\frac{t_1}{8\tilde{t}_2}\left(1-y^2\right)\right]+O\left(\frac{g^2}{t_1^2},\frac{g^2}{\tilde{t}_2^2}\right)\right\}.
\end{align}
We treat the $n=1$ band hopping energy $t_1$ as much smaller than the $n=2$ band hopping energy $\tilde{t}_2$, consistent with experimental values. Taylor expanding $g(V)$ to leading order in $(V-V_t)$ gives the two-body rate behavior near threshold with the leading order correction in $(V-V_t)/t_1$:
 \begin{align}
\Gamma(V\leq V_t)=&\Gamma_0\left[1-\left.\frac{7}{16}\frac{dg(V')}{dV'}\right\vert_{V_t}\left(\frac{V-V_t}{t_1}\right)\right],\\
\Gamma(V>V_t)=&0.
\end{align}
\subsection{Three-body decay rate as $V\rightarrow V_t^+$}
We now consider the three-body decay rate for well depths just above $V_t$. As $V\rightarrow V_t^+$, the three-body rate diverges because a virtual intermediate $n=2$ state becomes progressively closer to a real, energetically permissible state. The three-body rate is given by:
\begin{align}
\Gamma=&\frac{128\pi t_1^2}{\hbar U_{11}^2}\sum_{j\neq0}\int_{-1}^{1}\frac{dk_1}{2}\int_{-1}^{1}\frac{dk_4}{2}\cos^2\left(\pi k_1\right)\times\nonumber\\
&\ \ \ \left\vert\int_{-1}^{1}\frac{dk}{2}\frac{U_{11}^{02}(k)U_{12}^{04}(k,k_4)}{\epsilon_{1k}+\epsilon_1-\epsilon_0-\tilde{\epsilon}_{2k}}\right\vert^2\delta\left(\Delta E_{k_1k_4}\right),\\
\Delta E_{k_1k_4}=&\tilde{\epsilon}_{4k_4}+2\epsilon_0-2\epsilon_1-\epsilon_{1k_1}.
\end{align}
Near the singularity, the dominant contributions to the rate $\Gamma$ come from momenta near $(k,k_1)=(1,1)$. Pulling nonsingular terms like $U_{11}^{02}(1)$ outside the integral and exploiting left-right symmetry gives
\begin{subequations}
\label{array_threebod}
\begin{align}
 \Gamma\approx&2\Gamma_1\sum_{j>0}\int_{0}^{1}dk_1\left\vert\int_{-1}^{1}\frac{dk}{2}\frac{e^{ik\pi j}}{\epsilon_{1k_1}+\epsilon_1-\epsilon_0-\tilde{\epsilon}_{2k}}\right\vert^2,
\\
\Gamma_1\approx&\frac{128\pi t_1^2}{\hbar U_{11}^2}\frac{\left\vert U_{11}^{02}(1)\right\vert^2\left\vert U_{12}^{04}(1,k_4^0)\right\vert^2}{\left\vert d\tilde{\epsilon}_{k_4}/dk_4\vert_{k_4^0}\right\vert},
\end{align}
\end{subequations}
where $k_4^0$ is set by conservation of energy:
\begin{equation}
\tilde{\epsilon}_{4k_4^0}+2\epsilon_0-\epsilon_{1}-\epsilon_{11}=0.
\end{equation}
Using the parameter $g(V)$ from Eq.~(\ref{eq:g_def}), the $k-$integral in Eq.~(\ref{array_threebod}) can be performed using a contour:
\begin{subequations}
\begin{align}
&\int_{-1}^{1}\frac{dk}{2}\frac{e^{ik\pi j}}{\epsilon_{1k_1}+\epsilon_1-\epsilon_0-\tilde{\epsilon}_{2k}}=\frac{\left(\ell+\sqrt{\ell^2-1}\right)^j}{2t_2\sqrt{\ell^2-1}},\\
&\ell=-1+\frac{g-4t_1\cos^2\left(\frac{\pi k_1}{2}\right)}{2t_2}. 
\end{align}
\end{subequations}
Using the standard approximation $1-\alpha\approx\exp(-\alpha)$ for $\alpha\ll1$ and expanding $\ell^2-1$ to leading order in the small parameters $g/t_2$ and $t_1/t_2$, Eq.~(\ref{array_threebod}) becomes
\begin{equation}
\Gamma\approx\frac{\Gamma_1}{2t_2}\sum_{j>0}\int_{0}^{1}dk_1\frac{\exp\left[-2j\sqrt{\left(-g+4t_1\cos^2\frac{\pi k_1}{2}\right)/t_2}\right]}{-g+4t_1 \cos^2\frac{\pi k}{2}}.
\end{equation}
Summing the geometric series gives
\begin{equation}
\label{eqn:above_threshold} 
\Gamma\approx\frac{\Gamma_1}{32\sqrt{t_2t_1^3}}\sum_{j>0}\int_{0}^{1}dk_1\left(\frac{-g}{4t_1}+ \cos^2\frac{\pi k}{2}\right)^{-3/2}.
\end{equation}
If $g(V)/t_1<0$, as it is for $V>V_t$, this integral exists and is finite. For $V\leq V_t$ this integral diverges reflecting the presence of a real intermediate state. We can express the solution to Eq.~(\ref{eqn:above_threshold}) using the complete elliptic integral of the second kind with imaginary argument:
\begin{subequations}
\label{eq_elliptic_expans}
\begin{align}
 \int_{0}^{1}dx\left[K+\cos^2\left(\frac{\pi x}{2}\right)\right]^{-3/2}&=\frac{2}{\pi}\frac{\textrm{E}\left(i/\sqrt{K}\right)}{\sqrt{K}},\\
&=\frac{2}{\pi K}+O(\log K).
\end{align}
\end{subequations}
Substituting $-g/4t_1$ for $K$ in Eq.~(\ref{eq_elliptic_expans}) we have the final form for the three-body rate $\Gamma$ near the threshold value of well depth $V_t$:
\begin{equation}
 \Gamma\approx\frac{8\left(t_1\right)^{3/2}\left\vert U_{11}^{02}(1)\right\vert^2\left\vert U_{12}^{04}(1,k_4^0)\right\vert^2}{\hbar\sqrt{t_2}U_{11}^2\left\vert\frac{d\epsilon_{4k}}{dk}\right\vert_{k_4^0}\left[-\left.\frac{d}{dV'}g(V')\right\vert_{V_t}(V-V_t)\right]}.
\end{equation}

\section{\label{sec:bose_derive}Derivation of Boltzmann Equation with Bose Statistics}
In this Appendix, we derive the linearized Boltzmann equation at long times assuming Bose statistics. The principles of the derivation are nearly identical to those of Section \ref{subsec:classical}, but the formulas are notably more complicated. We work in the limit of a weak trap, where the semiclassical approximation is valid. We specify the occupation $f_{nkj}$ of band $n$, quasimomentum $k$ and site $j$ with the corresponding energy $\epsilon_{nkj}$ approximately given by:
\begin{align}
 \epsilon_{nkj}\approx\epsilon_{nk}+Aj^2+F_{nj},\\
 F_{nj}\equiv\frac{1}{N}\sum_{n'k'}U_{nn'}f_{n'k'j},
\end{align}
where $F_{nj}$ is the mean field shift. Since the $F_{nj}$ are small compared to the equilibrium temperature $T^e$ we can ignore the $j-$dependence and use $F_{nj}\approx F_{n0}$. The Bose distribution of $f_{nkj}$ at thermal equilibrium is given by: 
\begin{equation}
\label{eq:f_nkj_bose_eq}
f_{nkj}^e=N\left\{\exp\left[\left(\epsilon_{nkj}-\mu\right)/T^e\right]-1\right\}^{-1},
\end{equation}
where the chemical potential $\mu$ and equilibrium temperature $T^e$ are determined by simultaneously satisfying energy and number conservation:
\begin{subequations}
\begin{align}
\label{eq:en_cons_bose}
&\frac{1}{N}\sum_{nkj}\epsilon_{nkj}f_{nkj}=N\left(\epsilon_1+\frac{U_{00}}{3}\right),\\
&\frac{1}{N}\sum_{nkj}f_{nkj}=N.
\end{align}
\end{subequations}
To simplify the discussion we introduce $z_{nk}^e\equiv\exp\left[-\left(\epsilon_{nk}+F_{n0}-\mu\right)/T^e\right]$. Eq.~(\ref{eq:en_cons_bose}) simplifies to
\begin{equation}
\label{eq:en_cons_trap_bose}
\frac{\sum_{nk}\left(\epsilon_{nk}+F_{n0}\right)\textrm{Li}_{1/2}\left(z_{nk}^e\right)+\frac{T^e}{2}\textrm{Li}_{3/2}\left(z_{nk}^e\right)}{\sum_{nk}\textrm{Li}_{1/2}\left(z_{nk}^e\right)}=\epsilon_1+\frac{U_{00}}{3},
\end{equation}
where Li$_k$ is the polylogarithm; see Eq.~(\ref{eq:en_cons_trap}) for comparison. Neglecting spatial derivatives, the Boltzmann equation with these approximations becomes:
\begin{align}
\label{eq:boltz_full_v1_bose}
\frac{df_{n_1k_1j}}{dt}=-\sum_{n_2}\sum_{n_3\geq n_4}\frac{1}{N^3}\sum_{k_{2},k_3,k_4}
\Gamma_{n_1n_2j}^{n_3n_4}(k_1,k_2,k_3,k_4)\times\nonumber\\
\left[f_{n_1k_1j}f_{n_2k_2j}\left(1+f_{n_3k_3j}\right)\left(1+f_{n_4k_4j}\right)\right.\nonumber\\
\left.-f_{n_3k_3j}f_{n_4k_4j}\left(1+f_{n_1k_1j}\right)\left(1+f_{n_2k_2j}\right)\right],
\end{align}
with the function $\Gamma_{n_1n_2j}^{n_3n_4}$ given in Eq.~(\ref{eq:gamm_full}). We are most interested in scattering near the center of the trap, and so we set $j=0$ for the remainder of the discussion. We make the ansatz that within a given band the occupation $f_{nk0}$ is Bose distributed, but the relative occupation of bands $p_n$ can fluctuate. This ansatz is given by:
\begin{subequations}
\label{eq:ansatz}
\begin{align}
&f_{nk0}\approx\frac{\rho_n' p_n}{d_n(T)}\left\{\exp\left[\left(\epsilon_{nk}+F_{n0}-\mu\right)/T\right]-1\right\}^{-1},\\
&\rho_n'=\sqrt{\frac{4U_{00}}{\pi T^e}}\frac{\sum_k\left\{\exp\left[\left(\epsilon_{nk}+F_{n0}-\mu\right)/T^e\right]-1\right\}^{-1}}{\sum_k\textrm{Li}_{1/2}\left\{\exp\left[-\left(\epsilon_{nk}+F_{n0}-\mu\right)/T^e\right]\right\}},\\
&d_n(T)\equiv\frac{1}{N}\sum_k\left\{\exp\left[\left(\epsilon_{nk}+F_{n0}-\mu\right)/T\right]-1\right\}^{-1},
\end{align}
\end{subequations}
where the parameter $T$ is an effective temperature set by conservation of energy. At equilibrium, the band probabilities $p_n$ and mean field shifts $F_{n0}$ are given by:
\begin{subequations}
\begin{align}
 &p_n^e=d_n(T^e)/\rho_n',\\
&F_{n0}^e=\sum_{n'}U_{nn'}d_n(T^e).
\end{align}
\end{subequations}
Using the ansatz in Eq.~\ref{eq:ansatz} and summing Eq.~(\ref{eq:boltz_full_v1_bose}) over $k_1$ gives an equation for the $p_n$'s and $T$: 
\begin{widetext}
\begin{align}
\label{eq:boltz_full_bose}
&\frac{d}{dt}p_{n_1}=-\frac{1}{\rho_{n_1}'}\sum_{n_2,n_3\geq n_4}\frac{1}{N^4}\sum_{k_1k_2k_3k_4}\Gamma_{n_1n_20}^{n_3n_4}\left(k_{1},k_{2},k_{3},k_{4}\right)d_{n_1k_1}d_{n_2k_2}\left(1+d_{n_3k_3}\right)\left(1+d_{n_4k_4}\right)\times\nonumber\\
&\ \ \ \ \ \left[\frac{\rho_{n_1}'\rho_{n_2}'p_{n_1}p_{n_2}}{d_{n_1}d_{n_2}}\frac{\left(1+\frac{\rho_{n_3}'}{d_{n_3}}p_{n_3}d_{n_3k_3}\right)}{1+d_{n_3k_3}}\frac{\left(1+\frac{\rho_{n_4}'}{d_{n_4}}p_{n_4}d_{n_4k_4}\right)}{1+d_{n_4k_4}}-\frac{\rho_{n_3}'\rho_{n_4}'p_{n_3}p_{n_4}}{d_{n_3}d_{n_4}}\frac{\left(1+\frac{\rho_{n_1}'}{d_{n_1}}p_{n_1}d_{n_1k_1}\right)}{1+d_{n_1k_1}}\frac{\left(1+\frac{\rho_{n_2}'}{d_{n_2}}p_{n_2}d_{n_2k_2}\right)}{1+d_{n_2k_2}}\right],
\end{align}
\end{widetext}
where $d_{nk}$ is given by:
\begin{equation}
d_{nk}=\left\{\exp\left[\left(\epsilon_{nk}+F_{n0}-\mu\right)/T\right]-1\right\}^{-1}.
\end{equation}
Since the temperature is so much larger than the bandwidth of the relevant bands, we can approximate $d_{nk}$ by its band average $d_n$. We define $\gamma_{n_1n_2}^{n_3n_4}$ by
\begin{align}
\label{eq:gamma_nn2_n3n4_bose}
\gamma_{n_1n_2}^{n_3n_4}&=\nonumber\\
&\frac{1}{N^4}\sum_{k_{1-4}}\Gamma_{n_1n_20}^{n_3n_4}\left(k_{1-4}\right)d_{n_1}d_{n_2}\left(1+d_{n_3}\right)\left(1+d_{n_4}\right).
\end{align}
Expanding the Boltzmann equation [Eq.~(\ref{eq:boltz_full})] to first order in deviations in both $p_n$ and $T$ from equilibrium including these approximations gives
\begin{align}
\label{eqnarray:lin_boltz_eq_bose}
\frac{d\delta p_{n_1}}{dt}&=-\frac{1}{\rho_{n_1}'}\sum_{n_2,n_3\geq n_4}\gamma_{n_1n_2}^{n_3n_4}(T^e)\times\nonumber\\
&\bigg[{\cal P}_{n_1}\delta p_{n_1}+{\cal P}_{n_2}\delta p_{n_2}-{\cal P}_{n_3}\delta p_{n_3}-{\cal P}_{n_4}\delta p_{n_4}\nonumber\\
&\left.+\frac{\delta T}{T^e}\big(-\overline{{\cal E}}_{n_1}-\overline{{\cal E}}_{n_2}+\overline{{\cal E}}_{n_3}+\overline{{\cal E}}_{n_4}\big)\right.\nonumber\\
&\left.+\sum_m\delta p_m\left({\cal B}_{mn_1}+{\cal B}_{mn_2}-{\cal B}_{mn_3}-{\cal B}_{mn_4}\right)\right],
\end{align}
where the terms $\lbrace {\cal P}_n,\overline{{\cal E}_n},{\cal B}_{mn}\rbrace$ are defined by
\begin{subequations}
\begin{align}
&{\cal P}_n=\frac{\rho_{n}'}{d_{n}^e\left(1+d_{n}^e\right)},\\
&\overline{{\cal E}}_n=\frac{1}{T^ed_{n}^e\left(1+d_{n}^e\right)}\frac{1}{N}\sum_k\frac{\left(\epsilon_{nk}+F_{n0}^e-\mu\right)z_{nk}^e}{\left(1-z_{nk}^e\right)^2},\\
&{\cal B}_{mn}=\frac{\rho_m'U_{mn}}{T^ed_{n_1}^e\left(1+d_{n_1}^e\right)}\frac{1}{N}\sum_k\frac{z_{nk}^e}{\left(1-z_{nk}^e\right)^2},
\end{align}
\end{subequations}
with $d_n^e$ and $A_n^e$ are shorthands for $d_n(T^e)$ and $A_n(T^e)$ respectively. Since Eq.~(\ref{eq:boltz_full_bose}) does not mix $f_{nkj}$ on different sites we must have energy conservation on-site, giving the relation 
\begin{equation}
\label{eq:en_cons_site_bose}
\delta\sum_{nk}\left(\epsilon_{nk}+F_{n0}\right) f_{nk0}=0.
\end{equation}
Eq.~(\ref{eq:en_cons_site_bose}) leads to the following equation for $\delta T$ in terms of the $\delta p_n$:
\begin{equation}
\label{eq:delta_T_bose}
\delta T=\frac{-\left(T^e\right)^2\sum_{n}\rho_n'\delta p_n\left(\langle\epsilon_n\rangle+2F_{n0}^e\right)}{\sum_n\frac{1}{N}\sum_k\left(\epsilon_{nk}-\left\langle\epsilon_n\right\rangle\right)\left(\epsilon_{nk}+F_{n0}^e-\mu\right)\frac{z_{nk}^e}{\left(1-z_{nk}^e\right)^2}},
\end{equation}
with $\left\langle\epsilon_n\right\rangle$ given by:
\begin{equation}
 \left\langle\epsilon_n\right\rangle=\frac{1}{d_{n}^e}\frac{1}{N}\sum_kd_{nk}\epsilon_{nk}.
\end{equation}
We find the slowest-decaying eigenmode of Eq.~(\ref{eqnarray:lin_boltz_eq_bose}) in Section \ref{subsec:eigvec}. The decomposition of the slowest decaying mode $v_1$ on each of the $\delta p_n$ is given in Table \ref{tab:eigvector_bose}.
\bibliography{highbandmott_draft_final}
\end{document}